\documentclass[final,authoryear,3p,times,10pt]{elsarticle}

\usepackage{multirow,setspace,times,amssymb,amsmath,graphicx,color,rotating,subfigure,url}
\usepackage[colorlinks,linkcolor=blue,citecolor=blue]{hyperref}
\usepackage{booktabs}
\usepackage{multicol}
\usepackage{multirow}
\begin{document}

\begin{frontmatter}

\title{Multifractal characteristics and return predictability in the Chinese stock markets}
\author[DF]{Xin-Lan Fu \fnref{cofirstauthor}}
\author[DF]{Xing-Lu Gao \fnref{cofirstauthor}} 
\fntext[cofirstauthor]{These authors contribute equally.}
\author[DF]{Zheng Shan}
\author[DF]{Zhi-Qiang Jiang \corref{cor}} \ead{zqjiang@ecust.edu.cn} %
\author[DF,DM]{Wei-Xing Zhou \corref{cor}}
\cortext[cor]{Corresponding author. Address: 130 Meilong Road, P.O. Box 114, Department of Finance, East China University of Science and Technology, Shanghai 200237, China, Phone: +86 21 64250053, Fax: +86 21 64253152.}
\ead{wxzhou@ecust.edu.cn} %

\address[DF]{Department of Finance, School of Business, East China University of Science and Technology, Shanghai 200237, China}
%\address[RCE]{Research Center for Econophysics, East China University of Science and Technology, Shanghai 200237, China}
\address[DM]{Department of Mathematics, School of Science, East China University of Science and Technology, Shanghai 200237, China}

\begin{abstract}
By adopting Multifractal detrended fluctuation (MF-DFA) analysis methods, the multifractal nature is revealed in the high-frequency data of two typical indexes, the Shanghai Stock Exchange Composite 180 Index (SH180) and the Shenzhen Stock Exchange Composite Index (SZCI). The characteristics of the corresponding multifractal spectra are defined as a measurement of market volatility. It is found that there is a statistically significant relationship between the stock index returns and the spectral characteristics, which can be applied to forecast the future market return. The in-sample and out-of-sample tests on the return predictability of multifractal characteristics indicate the spectral width $\Delta {\alpha}$ is a significant and positive excess return predictor. Our results shed new lights on the application of multifractal nature in asset pricing.

\end{abstract}

\begin{keyword}
Multifractal Characteristics \sep Multifractal Detrended Fluctuation Analysis \sep Return Predictability
\end{keyword}

\end{frontmatter}

\section{Introduction}
\label{Sec:Introduction}
%\subsection{Background}

As we know, market returns do not have auto-correlations, but exhibit nonlinear long memory behaviors, which corresponding to the multifractal nature \citep{Muzy-Sornette-Delour-Arneodo-2001-QF, Calvet-Fisher-2002-RES, Jiang-Zhou-2011-PRE}. Two stylized facts in financial series, including fat-tailed distributions and long range dependence, are considered as the sources of multifractality \citep{Zhou-2009-EPL, Jiang-Zhou-2008a-PA, Zhou-2012-CSF, Grahovac-Leonenko-2014-CSF}. Multifractal nature in returns makes the price dynamics differ from the Brownian process, triggering the studies of applying multifractality on uncovering market efficiency \citep{Wang-Wu-2013-CE, Liu-Wang-Wan-2010-IRFA}, designing trading strategies \citep{Dewandaru-Masih-Bacha-Masih-2015-PA}, constructing measures for improving volatility forecasts \citep{Wei-Wang-2008-PA, Chen-Wu-2011-PA, Wei-Chen-Lin-2013-PA, Chen-Wei-Lang_Lin-Liu-2014-PA}, and to list a few. New theoretical models including multifractal random walk (MRW) \citep{Bacry-Delour-Muzy-2001-PRE}, the multifractal model of asset returns (MMAR) \citep{Calvet-Fisher-2002-RES}, and Markov-switching multifractal model (MSM) \citep{Calvet-Fisher-2001-JEm} have been proposed to replicate the multi-scaling price behaviors. In comparison to other competitive econometric models, these models can improve the performance of volatility forecasting \citep{Calvet-Fisher-2001-JEm, Calvet-Fisher-2004-JFEm, Duchon-Robert-Vargas-2012-MF, Chuang-Huang-Lin-2013-NAJEF, Lux-MoralesArias-Sattarhoff-2014-JFc, Nasr-Lux-Ajmi-Gupta-2016-IREF, Segnon-Lux-Gupta-2017-RSER, Wang-Wu-Li-2017-IJF}, predicting financial duration \citep{Chen-Diebold-Schorfheide-2013-JEm, Zikes-Barunik-Shenai-2017-EmR}, and estimating VaR \citep{Batten-Kinateder-Wagner-2014-PA, Lux-Kaizoji-2008-JEDC, Chuang-Huang-Lin-2013-NAJEF, Lee-Song-Chang-2016-PA, Lux-Segnon-Gupta-2016-EE, Herrera-Rodriguez-Pino-2017-EE}. The forecasting power of multifractality on volatility indicates that multifractal characteristics in price dynamics could have strong links to market risks. As high market risks are accompanied with high returns, we could infer that strong multifractality in price dynamics will have high returns. However, such inferences are still lack of empirical evidence.
 
The research framework of return predictability provides an avenue to uncover whether the multifractal characteristics can be employed as predictors to forecast returns. The return predictability has been received considerable research interests, because it can highlight the understanding of asset pricing in the academy and improve the performance of stock investments in the industry. However, the predictability of stock returns is under controversy. On one hand, \cite{Welch-Goyal-2008-RFS} perform both in-sample and out-of-sample tests on return predictability with the factors which are reported to be prominent from earlier academic research and found that almost all of the factors seem unstable or even spurious. On the other hand, recent researches reveal that the factors, including the unexpected changes of oil prices \citep{Casassus-Higuera-2012-QF}, cash flow volatility \citep{Narayan-Westerlund-2014-IRFA},  aligned technical indicator \citep{Lin-2017-JFinM}, the curvature of the oil futures curve \citep{Chiang-Hughen-2017-JBF}, price-to-fundamental ratios \citep{Lawrenz-Zorn-2017-JEF}, the daily internet search volume index (SVI) \citep{Chronopoulos-Papadimitriou-Vlastakis-2018-JIMF}, and to list a few, can significantly and economically predict the excess returns. Comparing with the market returns, the sectoral returns exhibit much stronger predictability. However, the forecasting ability of predictors will be destroyed after they are published \citep{Mclean-Pontiff-2016-JF}, which is evidenced by that the predictors exhibit time-varying predictability \citep{Devpura-Narayan-Sharma-2018-JIFMIM}.

Inspired by the potential connections between multifractal nature and market volatility, it would be interesting to test whether the multifractal characteristic could be a return predictor. Our work will fill this gap. This paper is organized as follows: Data and methods are given in Sec.~\ref{Sec:DataMethods}. Sec.~\ref{Sec:EmpMF} presents the results of empirical multifractality. The in-sample and out-of-sample tests on return predictability based on multifractality are elaborated in Sec.~\ref{Sec:PredictionMF}. Sec.~\ref{Sec:Conclusion} concludes.

\section{Data and Methods}
\label{Sec:DataMethods}
\subsection{Data sets}
Our data, including the Shanghai Stock Exchange Composite 180 Index (SH180) and the Shenzhen Stock Exchange Composite Index (SZCI) in the Chinese stock markets, are retrieved from the finance database of Resset (http://www.resset.cn). Both indexes cover a period from February 14, 2003 to December 31, 2015 including 3036 trading days in total. By removing the days having recording errors, we left 3024 days for SH180 and 3032 days for SZCI, respectively. There are four trading hours (240 minutes) on each trading day. For each index, we have the price $p_m$ at each minute on each trading day and thus we define the minutely return $r_m(t)$ as,
\begin{equation}
r_m(t) = \ln p_m(t) - \ln p_m(t-1).
\label{Eq:MFPredictReturn:MinReturn}
\end{equation}
We regard the last price on each trading day as the closing price $p_c(d)$ on that day and the daily return $r_d$ is defined in the following, 
\begin{equation}
r_d(d) = \ln p_c(d) - \ln p_c(d-1).
\label{Eq:MFPredictReturn:DReturn}
\end{equation}

\subsection{Multifractal detrended fluctuation analysis (MF-DFA)}

For a given window of minutely returns $r_m(i), i = 1, \cdots, N$, we can define $y(i)$ as follows,
\begin{equation}
 y(i) = \sum_{u=1}^{i} r_m(i),~i = 1, 2, \cdots, N.
  \label{Eq:cumsum}
\end{equation}
The series $y$ is covered by $N_s$ disjoint boxes and each box has
the same size $s$. For our convenience, we label the sub-series in
each box as,
\begin{equation}
 Y_k(i) = \{y(i)|(k-1)s+1 \leq i \leq ks\},
  \label{Eq:subseries}
\end{equation}
In some cases, the whole series $y$ cannot be exactly covered by
$N_s$ boxes, which means that we have to neglect some data points at
the end of the series. In order to avoid this situation, we can
utilize $2N_s$ boxes to cover the series, where $N_s$ boxes cover
from the beginning and $N_s$ boxes cover from the end. In each box,
the sub-series $Y_k$ is regressed by a polynomial $g_l(\cdot)$ of
order $l$ (in our work $l = 1$). The overall detrended fluctuation $F_q(s)$ of the sub-series $Y_k$ is 
defined via the sample variance of the fitting residuals as follows,
\begin{equation}
 F_q(s) = \left\{\frac{1}{2N_s}\sum_{k=1}^{2N_s}[F_k(s)]^q \right\}^{1/q}.
  \label{Eq:overallfluctuation}
\end{equation}
where $q$ can take any real value except for $q = 0$. While $q = 0$,
we have
\begin{equation}
 F_0(s) = \exp\left\{\frac{1}{2N_s}\sum_{k=1}^{2N_s}\ln[F_k(s)] \right\}.
  \label{Eq:overallfluctuation}
\end{equation}
according to l'H\^{o}pital's rule. By varying the value of $s$ in
the range from $s_{\min} = 20$ to $s_{\max} = N/4$, one can expect
the detrended fluctuation function $F_q(s)$ scales with the size
$s$, which reads
\begin{equation}
 F_q(s) \sim s^{h(q)},
  \label{Eq:scaling}
\end{equation}
where $h(q)$ is the generalized Hurst index. Note that while $q =
2$, $h(2)$ is nothing but Hurst index $H$. The scaling exponents
$\tau(q)$, which is used to reveal the multifractality in the
standard multifractal formalism based on partition function, can be
obtained from the following traditional function for each $q$,
\begin{equation}
 \tau(q) = qh(q) - \Delta f,
  \label{Eq:scalingfunction}
\end{equation}
where $\Delta f$ is the fractal dimension of the geometric support of the
multifractal measure (in our case $\Delta f = 1$). The local singularity
exponent $\alpha$ of the measure $\mu$ and its spectrum $f(\alpha)$
are related to $\tau(q)$ through the Legendre transformation
\citep{Halsey-Jensen-Kadanoff-Procaccia-Shraiman-1986-PRA},
\begin{equation} \label{Eq:alphaf}
\left\{ \begin{aligned}
         \alpha &= {\rm{d}}\tau(q)/{\rm{d}}q \\
                  f(\alpha)&=q \alpha -\tau(q)
                  \end{aligned} \right.~.
                          \end{equation}
Taking into account the statistical significance of the estimation
of overall fluctuation functions, we focus on $q \in [-4, 8]$.

We further employ these three parameters ($\Delta {\alpha}$,  $\Delta f$, and  $B$) to capture the overall characteristics of the multifractal spectrum. 
Parameter $\Delta \alpha$ stands for the width of multifractal spectrum, defined as $\Delta {\alpha} = \alpha_{\max} - \alpha_{\min}$. 
$\Delta {\alpha}$ quantitatively describes the dispersion of singularity exponents $\alpha$, and thus measures the degree of heterogeneity for the probability measure of subsets on the overall fractal structure according to the definition of $\alpha$ \citep{Zhou-2007}. In practice, $\Delta {\alpha}$ is widely used to gauge the degree of multifractality \citep{Jiang-Zhou-2008a-PA, Jiang-Zhou-2008b-PA, Zhou-2009-EPL}. The larger the value of $\Delta {\alpha}$ is, the stronger the multifractal nature is. 
Parameter $\Delta f$ is estimated via $\Delta f = f(\alpha_{\min}) - f(\alpha_{\min})$, depicting the difference between the proportion of the subset with the minimum probability measure and that with the maximum probability measure. Thus, more measures are at the peak if $\Delta f < 0$ and more measures are at the trough if $\Delta f > 0$. Parameter $B$ is obtained by fitting the $f(\alpha) \sim \alpha$ curve to the following quadratic function $f(\alpha) = A(\alpha - \alpha_0)^2 + B(\alpha - \alpha_0) +C$ \citep{Shimizu-Thurner-Ehrenberger-2002-Fractals, MunozDiosdado-RioCorrea-2006-IEEEconf}. $B$ indicates the asymmetry of the spectrum. If the absolute value of $B$ approaches to 0, the spectrum curve is more favorable to being symmetric. $B < 0$ means that the spectrum curve is right-hooked, indicating that the data set is dominated by the subsets with large probability measures. In contrast, $B > 0$ means that the spectrum exhibit a left-hooked pattern, suggesting that the subsets with small probability measures take the leading role in the data set. 

The shape of the multifractal spectrum and the definition of $\Delta f$ and $B$ underlie a positive correlation between $\Delta f$ and $B$. When $B > 0$ (respectively, $B < 0$), the multifractal spectrum is left-hooked (respectively, right-hooked), which means $f(\alpha_{\min}) > f(\alpha_{\max})$ (respectively, $f(\alpha_{\min}) < f(\alpha_{\max})$), and then we have $\Delta f > 0$ (respectively, $\Delta f < 0$). When $B$ approaches 0, the multifractal spectrum gradually becomes symmetric, which leads to $f(\alpha_{\min}) \approx f(\alpha_{\max})$ and we will obtain $\Delta f \approx 0$. However, $\Delta f$ and $B$ have very different physical meaning. $\Delta f$ measures the proportion difference between the minimum measure and the maximum measure, while $B$ indicates which type of measures plays an important role in the data sets, large measures or small measures.

\section{Empirical multifractal characteristics}
\label{Sec:EmpMF}

\begin{figure}[htb]
\small
\centering
\vskip 2.5mm
\includegraphics[width=5cm]{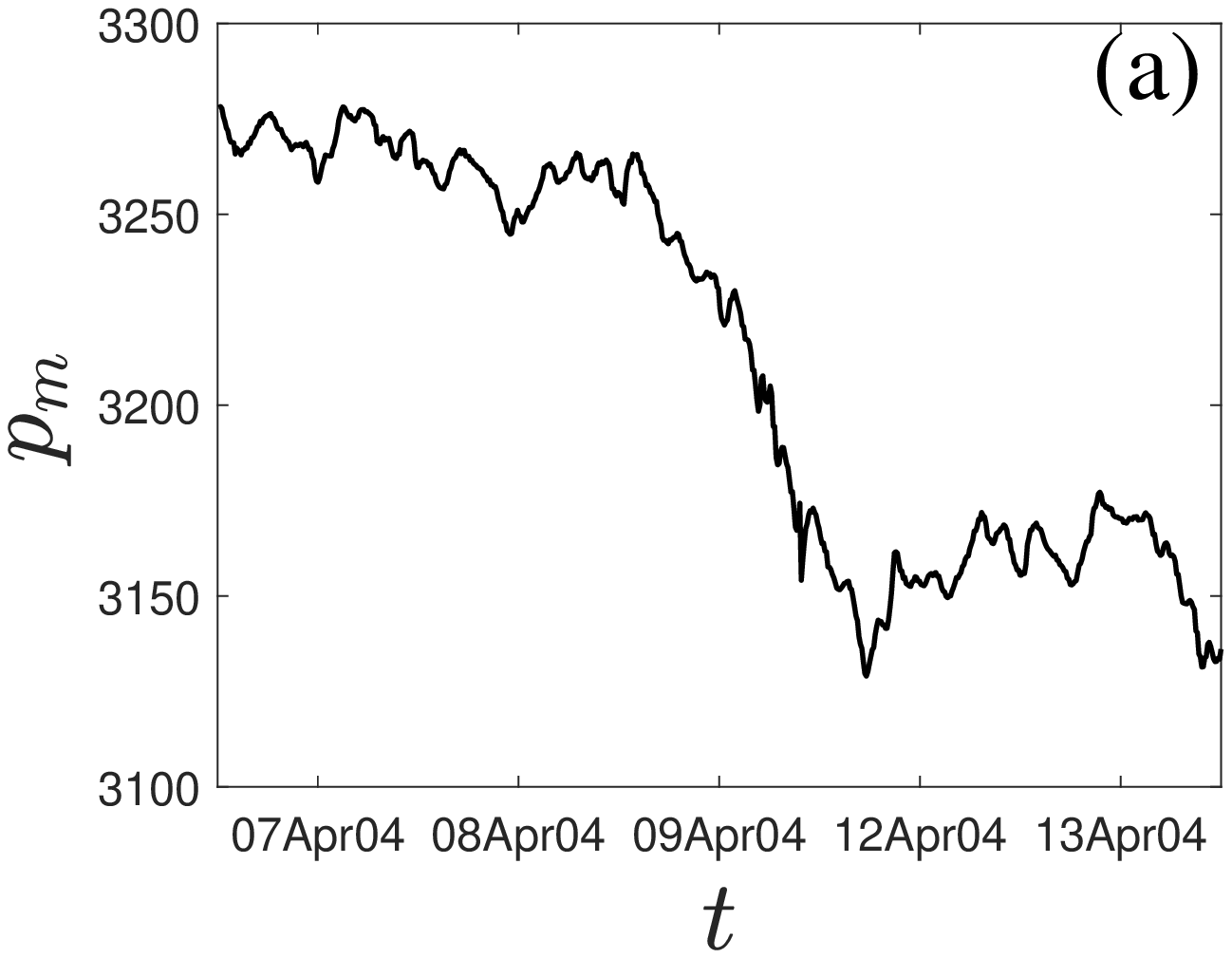} \hskip 2mm
\includegraphics[width=5.1cm]{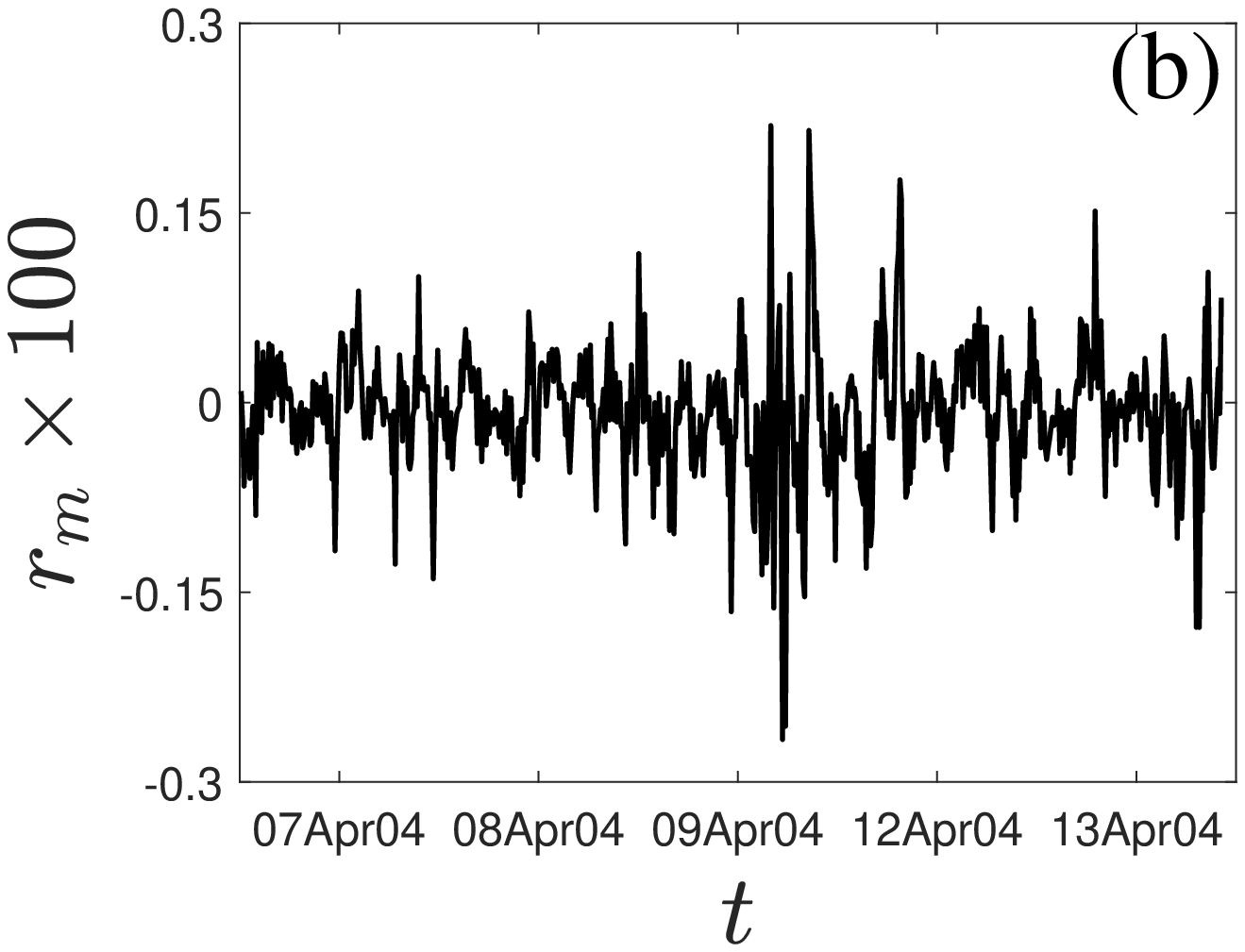} \hskip 2mm
\includegraphics[width=5.35cm]{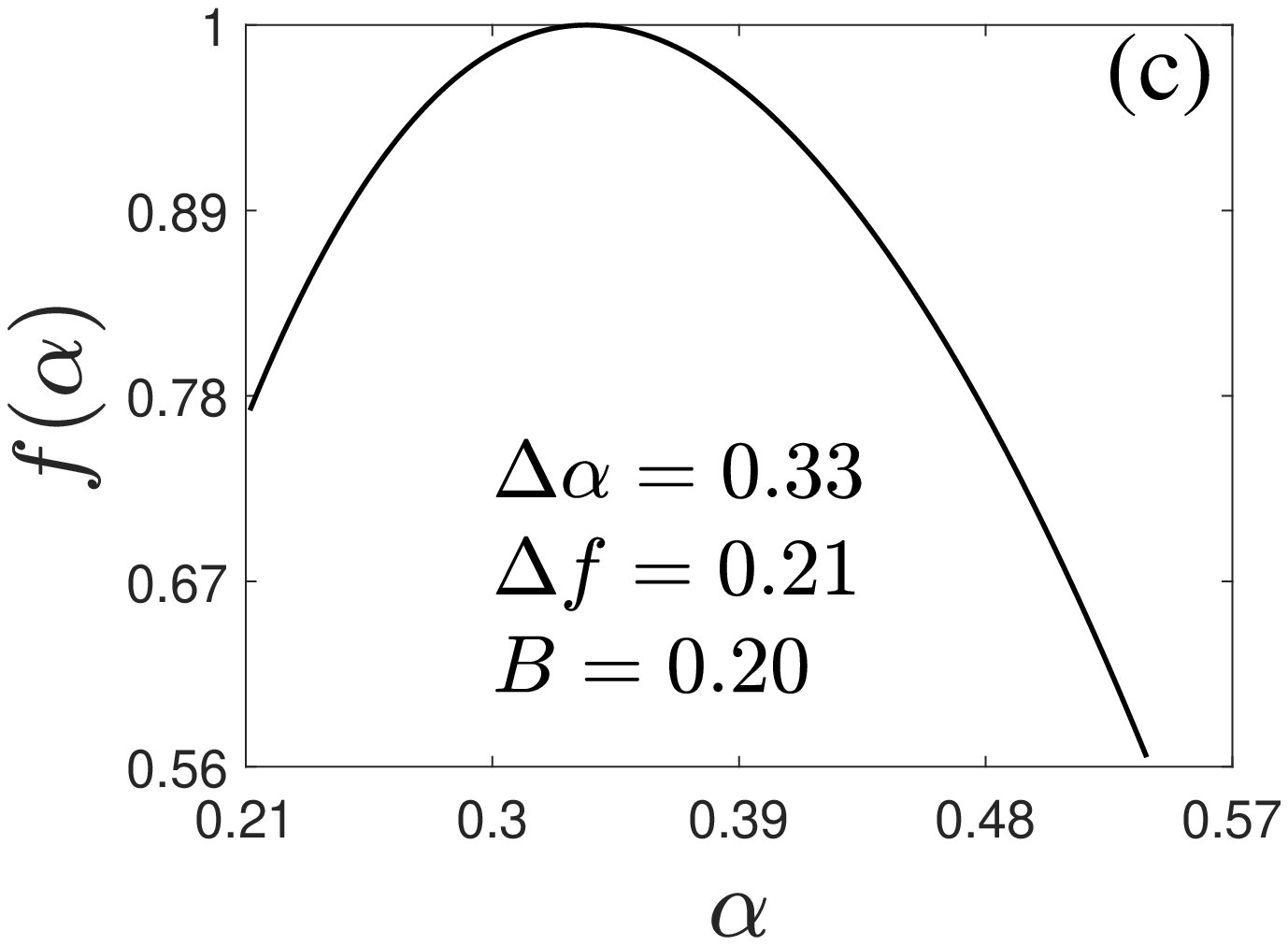}\hskip 2mm 
\vskip 2.5mm
\includegraphics[width=5cm]{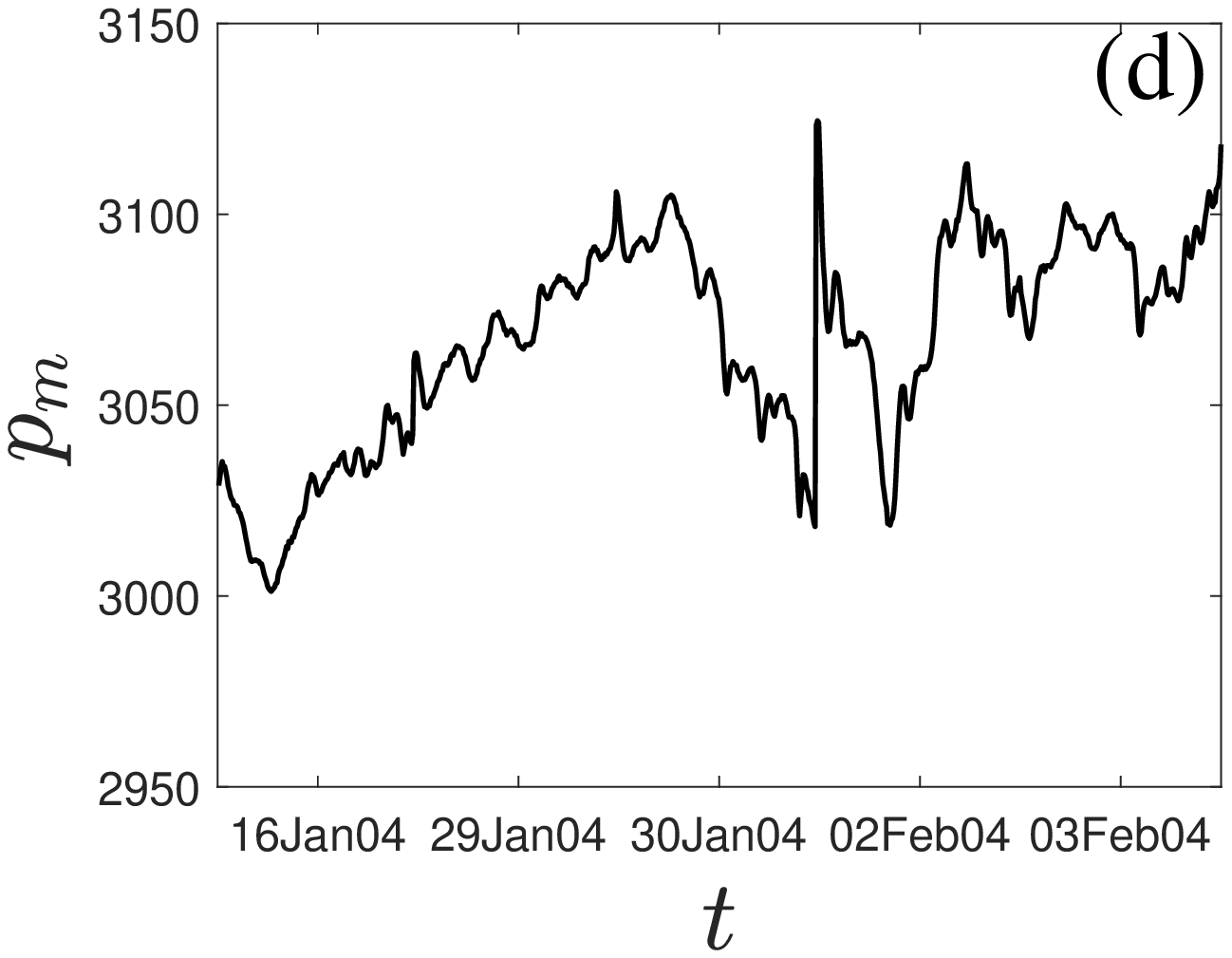} \hskip 2mm
\includegraphics[width=5.1cm]{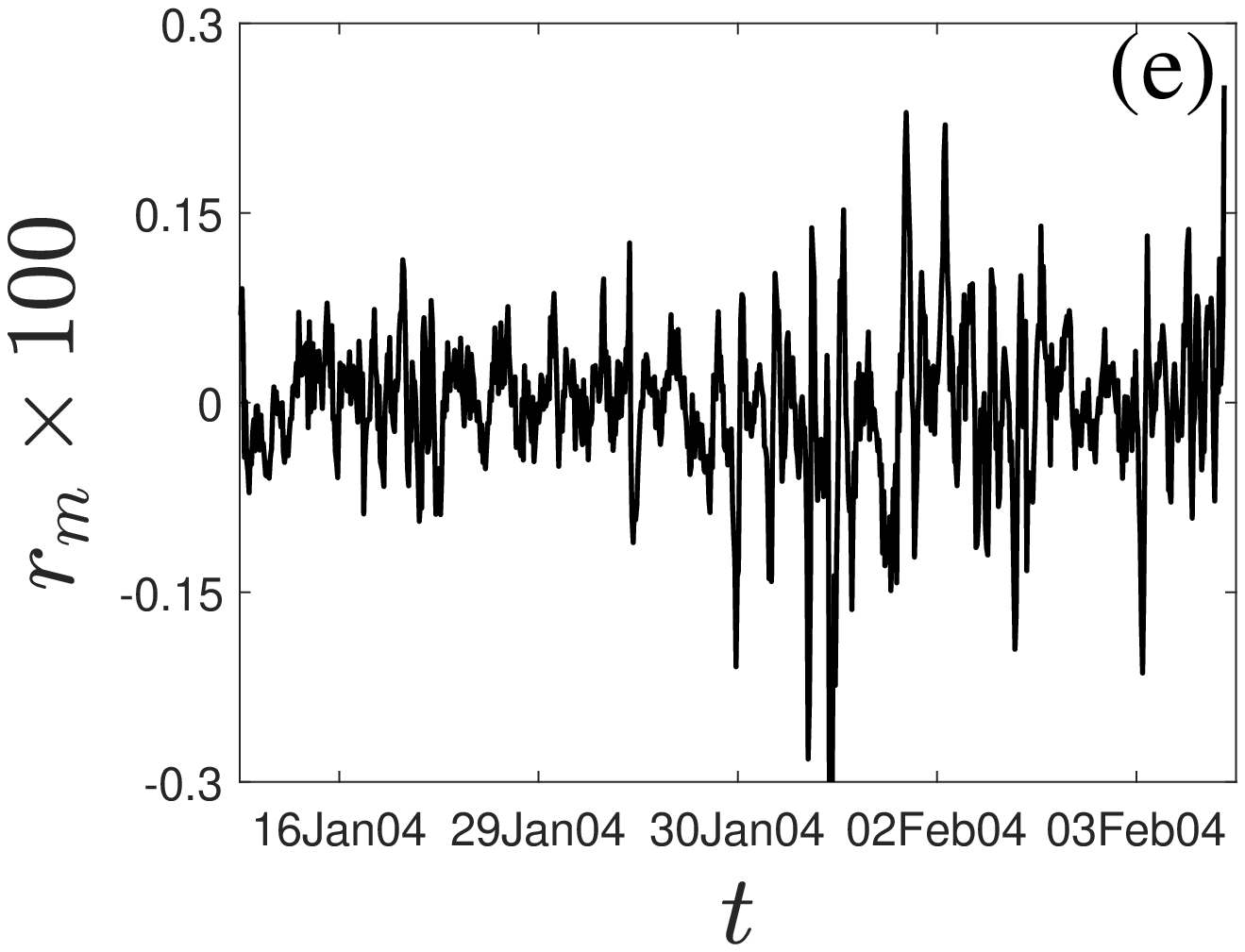} \hskip 2mm
\includegraphics[width=5.35cm]{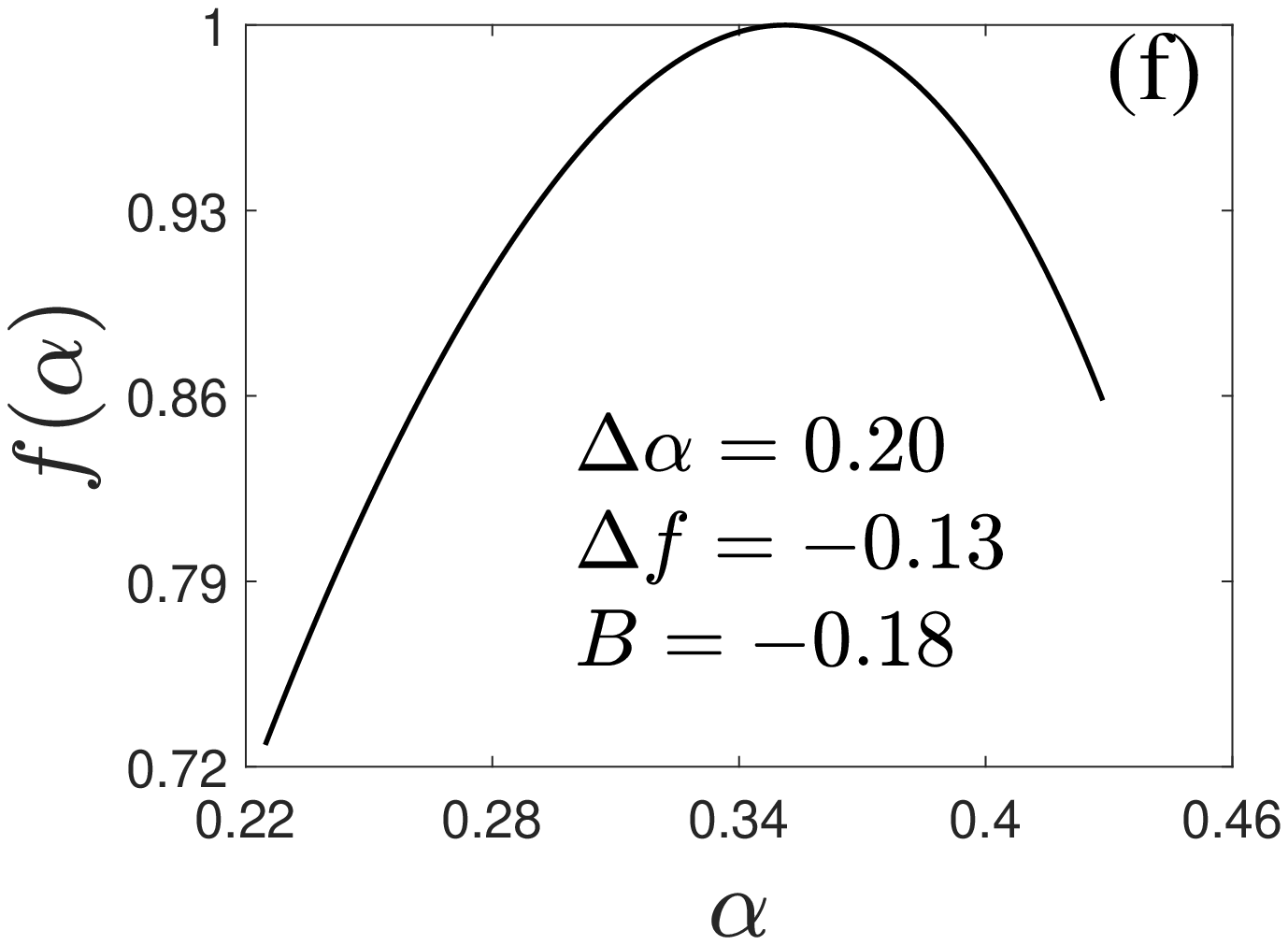} \hskip 2mm
\vskip 2.5mm
\includegraphics[width=5cm]{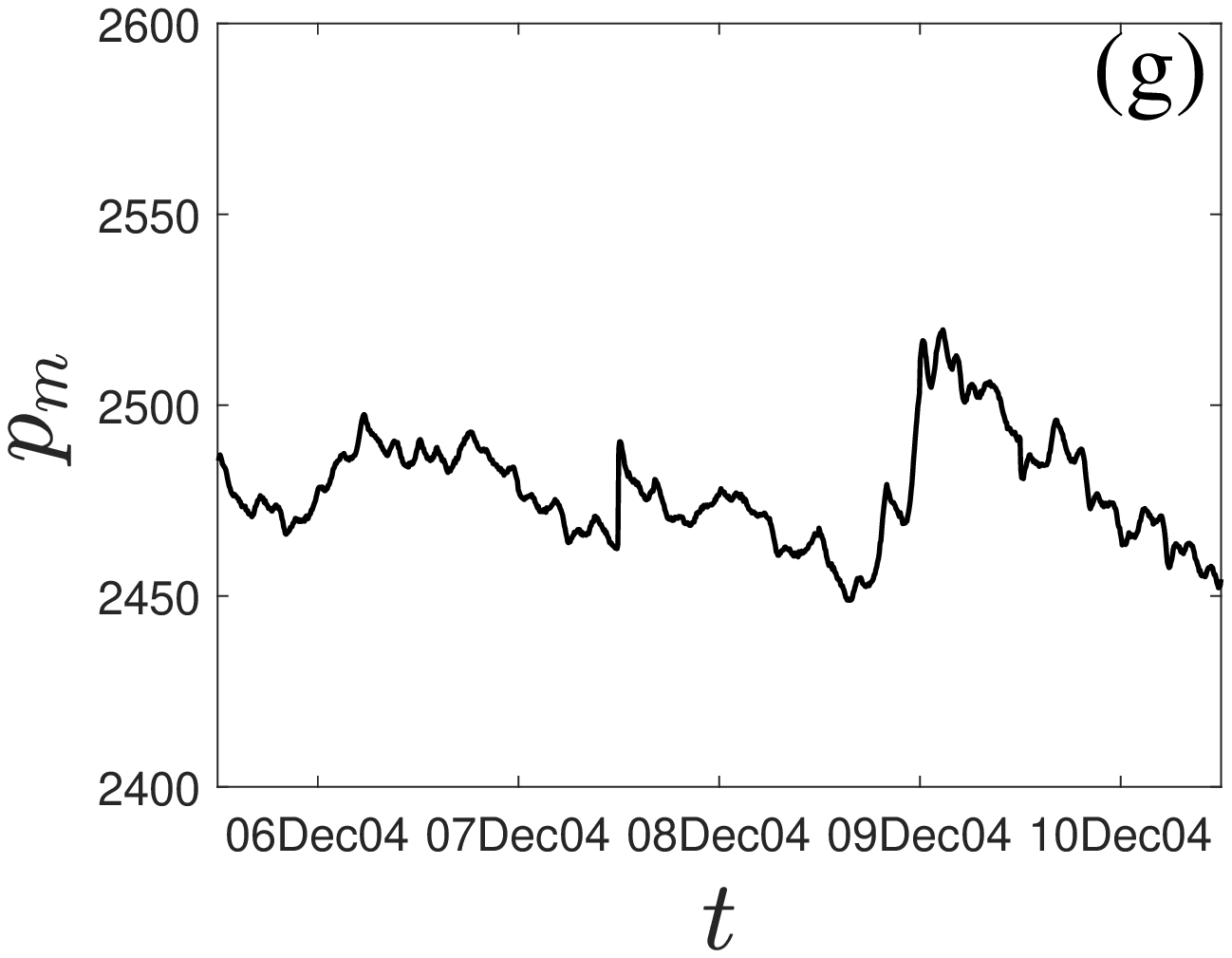} \hskip 2mm
\includegraphics[width=5.1cm]{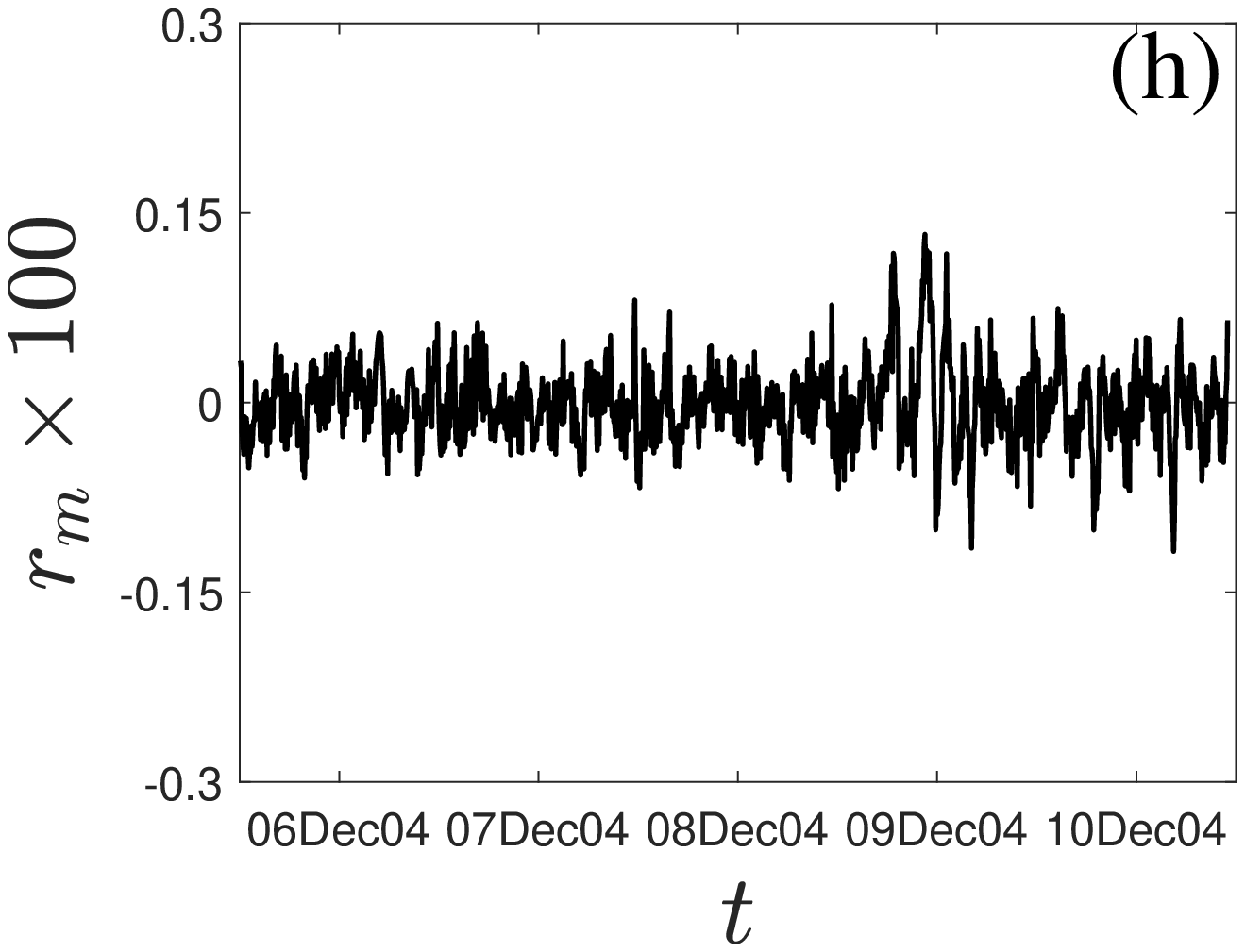} \hskip 2mm
\includegraphics[width=5.35cm]{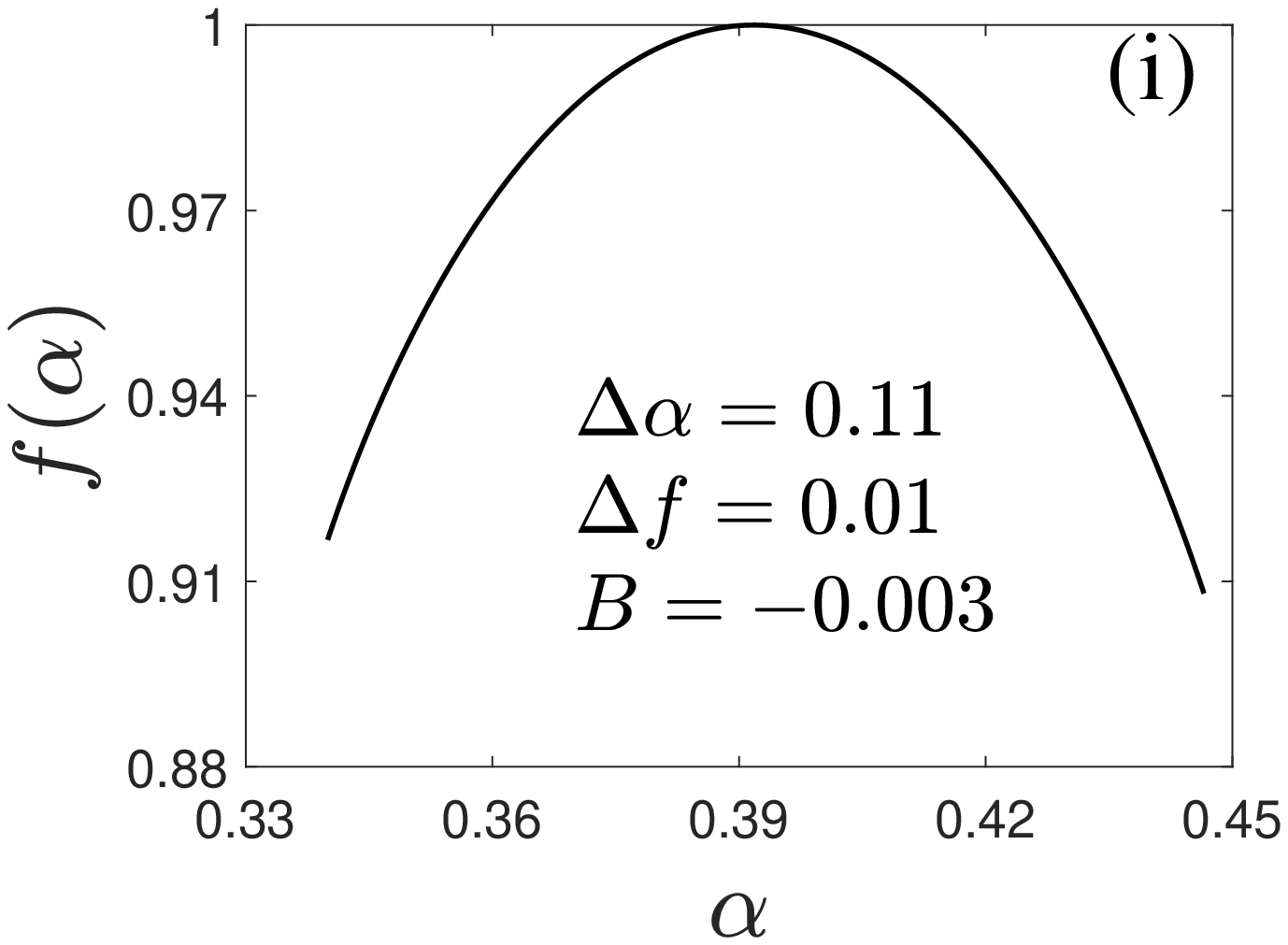} \hskip 2mm
\caption{\label{Fig:MF:HFReturns:3Windows} Results of multifractal analysis on SH180 in three windows, in which the prices exhibit a declining, rising, and sideways trend, respectively.  (a $-$c) Results of window 1, spanning from 7 April 2004 to 13 April 2004. (d $-$f) Results of window 2, spanning from 16 January 2004 to 3 February 2004. (g $-$i) Results of window 3, spanning from 6 December 2004 to 10 December 2004. (a, d, g) Plots of the price trajectories. (b, e, h) Plots of return series. (c, f, i) Plots of the multifractal spectrum. }
\end{figure}

Using a moving window with a size of 5 days, we perform the multifractal analysis on the returns in each window by means of the MFDFA method. To have an impression that the multifractal spectrum is able to quantitatively capture the market dynamics, we present the results of multifractal analysis on three typical price trajectories, corresponding to declining, rising, and sideways trends, in three windows for SH 180. Window 1 covers a period from 7 April 2004 to 13 April 2004. As shown in Fig.~\ref{Fig:MF:HFReturns:3Windows} (a), (b), and (c), in window 1, the price exhibits an obvious decreasing pattern and we will observe more small values in returns, which directly leads to a left-hooked multifractal spectrum. Window 2 spans a period from 16 January 2004 to 3 February 2004 and in that window the price has an increasing pattern and the returns contain more large values, which results in a right-hooked multifractal spectrum, as illustrated in Fig.~\ref{Fig:MF:HFReturns:3Windows} (d), (e), and (f). Window 3 is from 6 December 2004 to 10 December 2004. The results of window 3 plotted in Fig.~\ref{Fig:MF:HFReturns:3Windows} (g), (h), and (i) demonstrate that a symmetric multifractal spectrum results from the fact that the fraction of large and small returns are almost equal, especially for the case that the market price is in a sideways trend.

We also estimate the three characteristic parameters $(\Delta {\alpha}, \Delta f, B)$ of multifractal spectrum in the three windows and obtain $\Delta {\alpha} = 0.33$, $\Delta f = 0.21$, and $B = 0.20$ for window 1,  $\Delta {\alpha} = 0.20$, $\Delta f = -0.13$, and $B = -0.18$ for window 2,  $\Delta {\alpha} = 0.11$, $\Delta f = 0.01$, and $B = -0.003$ for window 3, respectively. For the width of multifractal spectra, we have $\Delta {\alpha}^{\rm{Window1}} > \Delta {\alpha}^{\rm{Window2}} > \Delta {\alpha}^{\rm{Window3}}$. This result is consistent with the observation that the market in window 1 is the most volatile and the market in window 3 is the least volatile, indicating that the spectral width $\Delta {\alpha}$ can be used to quantitatively capture the market volatility. \cite{Wei-Wang-2008-PA} have proposed a volatility measure based on the width of multifractal spectrum and such multifractal volatility is able to provide better VaR measures comparing with the GARCH-type models \citep{Wei-Chen-Lin-2013-PA}. We also find that the values of $\Delta f$ and $B$ in the three windows are in consistence with the geometric features of their multifractal spectra. 

By performing the multifractal analysis on the return series in each moving window, we will accumulate three series of the multifractal characteristics ($\Delta {\alpha}$, $\Delta f$, and $B$). Table~\ref{Tb:RI:Statistics} lists the basic information of the cumulative return ($\sum r_m$) and the multifractal characteristics ($\Delta {\alpha}$, $\Delta f$, and $B$) for SSE Constituent Index and SZSE Component Index. In Panel A, one can see that the skewness is positive and the kurtosis is much greater than 3 for $\sum r_m$, $\Delta {\alpha}$, $\Delta f$, and $B$, indicating that the four parameters all exhibit a right-skewed and fat-tailed distribution. It is also observed that for $\Delta {\alpha}$ and $\Delta f$ the gap between the mean and the median are very small and for $B$ the mean is less than the median, indicating that the extreme values in $\Delta {\alpha}$, $\Delta f$, and $B$ have negligible effects on the mean values. 

In Panel B of Table~\ref{Tb:RI:Statistics}, one can find that there is no correlation between the daily returns following moving windows $r_d$ and the three multifractal parameters ($\Delta {\alpha}$, $\Delta f$, and $B$), as their correlation coefficients are very close to 0 and none of them is significant. The cumulative returns in moving windows $\sum r_m$ are found to be negatively and weakly correlated with $\Delta {\alpha}$ for SH180 and significantly correlated with $\Delta f$ for both indexes. The correlation of $\Delta {\alpha}$ and $\Delta f$ is weak, positive, and significant for both indexes, because both coefficients are around 0.05. We also see that $\Delta f$ and $B$ are significantly positive correlated and the correlation coefficients are 0.14 and 0.13 for SH180 and SZCI, consisting with the underlying correlation between $\Delta f$ and $B$ according to their definitions and the multifractal spectrum. 

In Panel C of Table~\ref{Tb:RI:Statistics}, we find that $\Delta {\alpha}$ and $\Delta f$ exhibit very strong autocorrelated behaviors, since their autocorrelation coefficients of lags 1 and 5 are positive, large, and significant. The autocorrelations of cumulative returns $\sum r_m$ in moving windows are around 0.78 at lag 1 for both indexes and quickly fall to 0 with the increasing of lag, as the autocorrelations of lag 5 are around 0. The autocorrelation in $B$ is weak, because only significantly positive coefficients at lag 1 are observed and the values are less 0.1 for both indexes. In addition, the Ljung and Box Q statistics of lags 30 and 50 are statistically significant for the four parameters, showing that the null hypothesis of no autocorrelation up to the 10th and 20th orders is rejected at the 1\% level. Such results imply the existence of autocorrelation in $\sum r_m$, $\Delta {\alpha}$, $\Delta f$, and $B$, which can be attributed to that all the four parameters are estimated from overlapping moving windows. 

We report the results of augmented Dickey-Fuller (ADF) unit root tests and ARCH tests in Panels D and E of Table~\ref{Tb:RI:Statistics}. For the ADF unit root test, the optimal lag length is determined according to the Schwarz information criterion. For both indexes, all the ADF statistics show the rejection of the null hypothesis of a unit root in the series of $\sum r_m$, $\Delta {\alpha}$, $\Delta f$, and $B$ at the 1\% level, indicating the stationary of the four series. Panel F lists the $F$-statistics of the ARCH tests at lags 1, 5, 10 and 15, suggesting that the null hypothesis of no ARCH effects is rejected at any level of significance. The results clearly speak for the presence of heteroscedasticity in the series of $\sum r_m$, $\Delta {\alpha}$, $\Delta f$, and $B$ for both indexes.

\begin{table}[htb]
\setlength\tabcolsep{1.6pt}
%\footnotesize
\centering %\small
 \caption{\label{Tb:RI:Statistics} Basic information of the cumulative return ($\sum r_m$) and multifractal characteristics ($\Delta {\alpha}$, $\Delta f$, and $B$) in moving windows for SH180 and SZCI. Panel A reports mean, median, maximum (max), minimum (min), standard deviation (stdev), skewness (skew), and kurtosis (kurt) of the four parameters. Panel B lists the correlation coefficients between pairs of following variables,  $r_d$ (daily returns following moving windows), $\sum r_m$,  $\Delta {\alpha}$, $\Delta f$, and $B$ in moving windows. The autocorrelation coefficients of lag 1 and 5 and the Ljung-Box Q tests of lag 30 and 50 are presented in Panel C. The augmented Dickey-Fuller (ADF) unit root tests and ARCH tests are listed in Panels D and E. }
 \medskip
 \centering
\begin{tabular}{lcr@{.}lr@{.}lr@{.}lr@{.}lcr@{.}lr@{.}lr@{.}lr@{.}l}
\toprule
&& \multicolumn{8}{c}{SH180} && \multicolumn{8}{c}{SZCI}\\  %
 \cline{3-10} \cline{12-19}
   &&  \multicolumn{2}{c}{$\sum r_m$} & \multicolumn{2}{c}{$\Delta {\alpha}$} & \multicolumn{2}{c}{$\Delta f$}  & \multicolumn{2}{c}{$B$}  &&  \multicolumn{2}{c}{$\sum r_m$}  &  \multicolumn{2}{c}{$\Delta {\alpha}$} & \multicolumn{2}{c}{$\Delta f$}  & \multicolumn{2}{c}{$B$} \\
\midrule
\multicolumn{19}{l}{Panel A: Basic statistics} \\
 mean && $-$0&0025 & 0&1709& 0&0431&0&0520 && $-$0&0029 & 0&1533 & 0&0217 & 0&0308 \\
 median && $-$0&0024 & 0&1639 & 0&0408 & 0&0751 && $-$0&0032 & 0&1479 & 0&0237 & 0&0460\\
 max && 0&2380 & 0&5268 & 0&7315 & 7&3194 && 0&2620 & 0&6779 & 0&7756 & 7&1260\\
 min && $-$0&1904 & 0&0046 & $-$0&4040 & $-$4&6075 && $-$0&1703 & 0&0131 & $-$0&4234 & $-$4&7179\\
 stdev && 0&0408 & 0&0735 & 0&1083 & 0&4596 && 0&0440 & 0&0651 & 0&1002 & 0&4510\\
 skew && 0&2732 & 0&5311 & 0&2393 & 4&2459 && 0&3749 & 1&1539 & 0&4752 & 5&5109\\
 kurt && 5&0878 & 3&2767 & 5&8974 & 114&6074 && 5&1531 & 8&2139 & 8&4411 & 117&4338\\

\midrule
\multicolumn{19}{l}{Panel B: Correlations} \\
 $r_d$ && \multicolumn{2}{l}{} & 0&0013 & 0&0041 & 0&0093 && \multicolumn{2}{l}{} & 0&0127 & 0&0084 & 0&0206 \\
             && \multicolumn{2}{l}{} & (0&9437) & (0&8223) & (0&6079) && \multicolumn{2}{l}{} & (0&4862) & (0&6439) & (0&2568) \\
 $\sum r_m$ && \multicolumn{2}{l}{} & $-$0&0720 & 0&0443 & 0&0135 && \multicolumn{2}{l}{} & $-$0&0229 & 0&0641 & 0&0086 \\
             && \multicolumn{2}{l}{} & (0&0001) & (0&0148) & (0&4581) && \multicolumn{2}{l}{} & (0&2068) & (0&0004) & (0&6361) \\
 $\Delta {\alpha}$ && \multicolumn{4}{l}{} & 0&0678 & $-$0&0046 &&  \multicolumn{4}{l}{}  & 0&0492 & $-$0&0221 \\
              && \multicolumn{4}{l}{} & (0&0002) & (0&7983) &&  \multicolumn{4}{l}{}  & (0&0067) & (0&2243) \\
 $\Delta f$ &&   \multicolumn{6}{l}{}  & 0&1424 &&  \multicolumn{6}{l}{}  & 0&1302 \\
             && \multicolumn{6}{l}{}  & (0&0000) &&  \multicolumn{6}{l}{}  & (0&0000) \\

 \midrule
\multicolumn{19}{l}{Panel C: Autocorrelations} \\
rho(1) && 0&7768 & 0&8310 & 0&6991 & 0&0664 && 0&7897 & 0&7806 & 0&6940 & 0&0908 \\
             && (0&0000) & (0&0000) & (0&0000) & (0&0003) && (0&0000) & (0&0000) & (0&0000) & (0&0000) \\
rho(5) && 0&0230 & 0&5408 & 0&2674 & $-$0&0232 && 0&0508 & 0&3359 & 0&2580 & 0&0286 \\
              && (0&2054) & (0&0000) & (0&0000) & (0&2014) && (0&0051) & (0&0000) & (0&0000) & (0&1156) \\
Q(30) && 3613&8179 & 25486&8782 & 6554&7762 & 149&9445 && 3703&7547 & 11416&3511 & 6347&6620 & 209&2306 \\
             && (0&0000) & (0&0000) & (0&0000) & (0&0000) && (0&0000) & (0&0000) & (0&0000) & (0&0000) \\
Q(50) && 3844&0405 & 37193&0242 & 8025&5737 & 216&6295 && 4018&3729 & 15567&3894 & 7442&4816 & 233&5369 \\
             && (0&0000) & (0&0000) & (0&0000) & (0&0000) && (0&0000) & (0&0000) & (0&0000) & (0&0000) \\

 \midrule
\multicolumn{19}{l}{Panel D:  Unit root tests} \\
Optimal lag && 5& & 8& & 6& & 1& && 8& & 7& & 7& & 0& \\
ADF Statistics  && $-$15&5558 & $-$2&7288 & $-$10&8907 & $-$39&7719 && $-$16&0267 & $-$3&3889 & $-$11&0533 & $-$50&0434 \\
      && (0&0010) & (0&0066) & (0&0010) & (0&0010) && (0&0010) & (0&0010) & (0&0010) & (0&0010) \\

 \midrule
\multicolumn{19}{l}{Panel E:  ARCH tests} \\
ARCH(1) && 1177&5677 & 1291&7485 & 985&2632 & 119&4355 && 1302&2498 & 1821&5172 & 1262&2426 & 185&2461 \\
              && (0&0000) & (0&0000) & (0&0000) & (0&0000) && (0&0000) & (0&0000) & (0&0000) & (0&0000) \\
ARCH(5) && 1187&2470 & 1308&7620 & 997&9737 & 143&0733 && 1308&9308 & 1851&4963 & 1323&5307 & 200&2927 \\
              && (0&0000) & (0&0000) & (0&0000) & (0&0000) && (0&0000) & (0&0000) & (0&0000) & (0&0000) \\
ARCH(10) && 1189&5761 & 1334&9516 & 1007&7098 & 143&9011 && 1312&2320 & 1883&9523 & 1334&4502 & 290&5531 \\
              && (0&0000) & (0&0000) & (0&0000) & (0&0000) && (0&0000) & (0&0000) & (0&0000) & (0&0000) \\
ARCH(15) && 1191&2654 & 1343&3768 & 1009&9590 & 145&2063 && 1326&4128 & 1887&0393 & 1339&9717 & 299&9928 \\
              && (0&0000) & (0&0000) & (0&0000) & (0&0000) && (0&0000) & (0&0000) & (0&0000) & (0&0000) \\

\bottomrule
\end{tabular}
\end{table}

\section{Predictive power of multifractal characteristics}
\label{Sec:PredictionMF}

\subsection{In-sample tests}

\begin{figure}[htb]
\small
\centering
\vskip 2.5mm
\includegraphics[width=5cm]{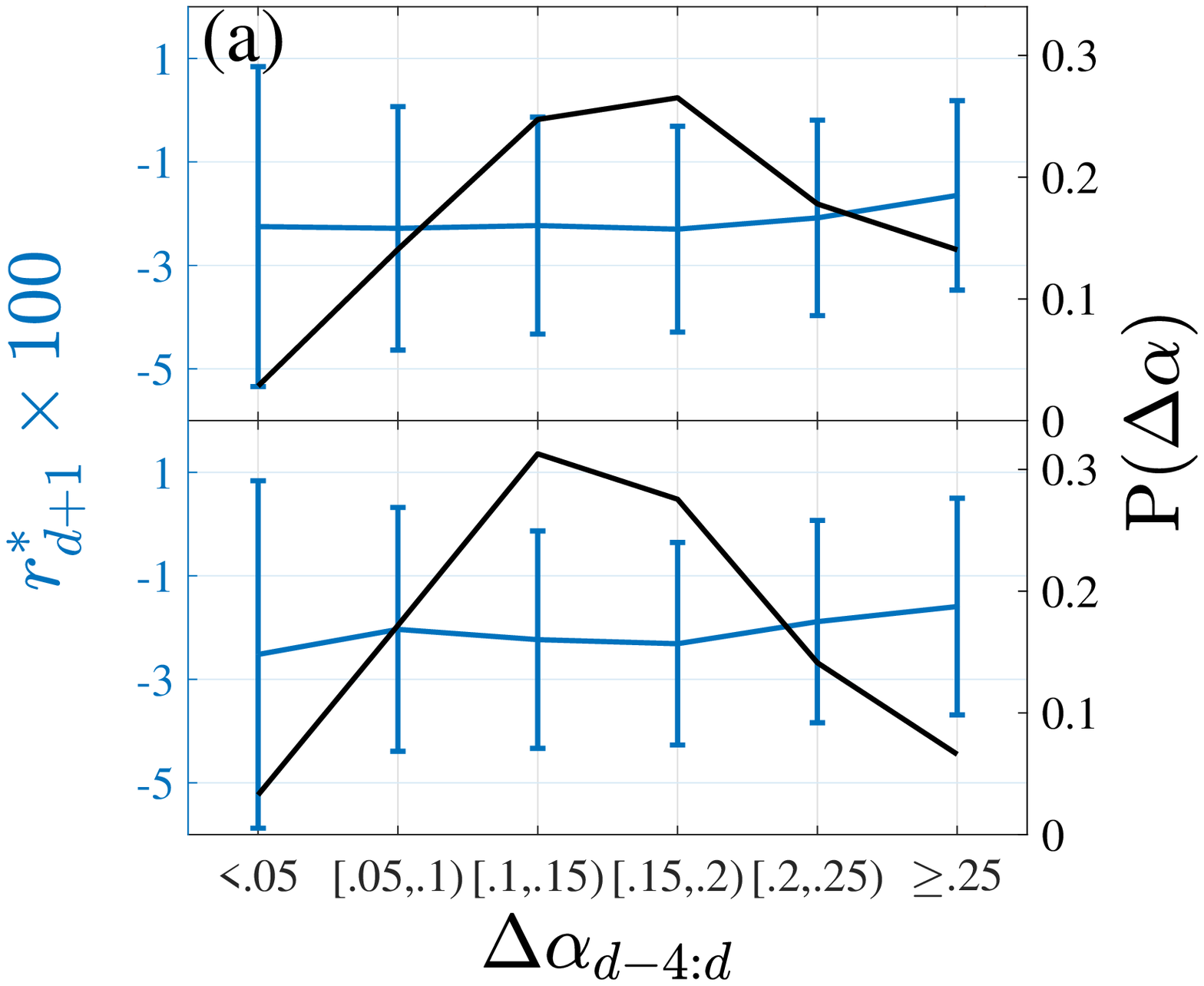} \hskip 3mm
\includegraphics[width=5cm]{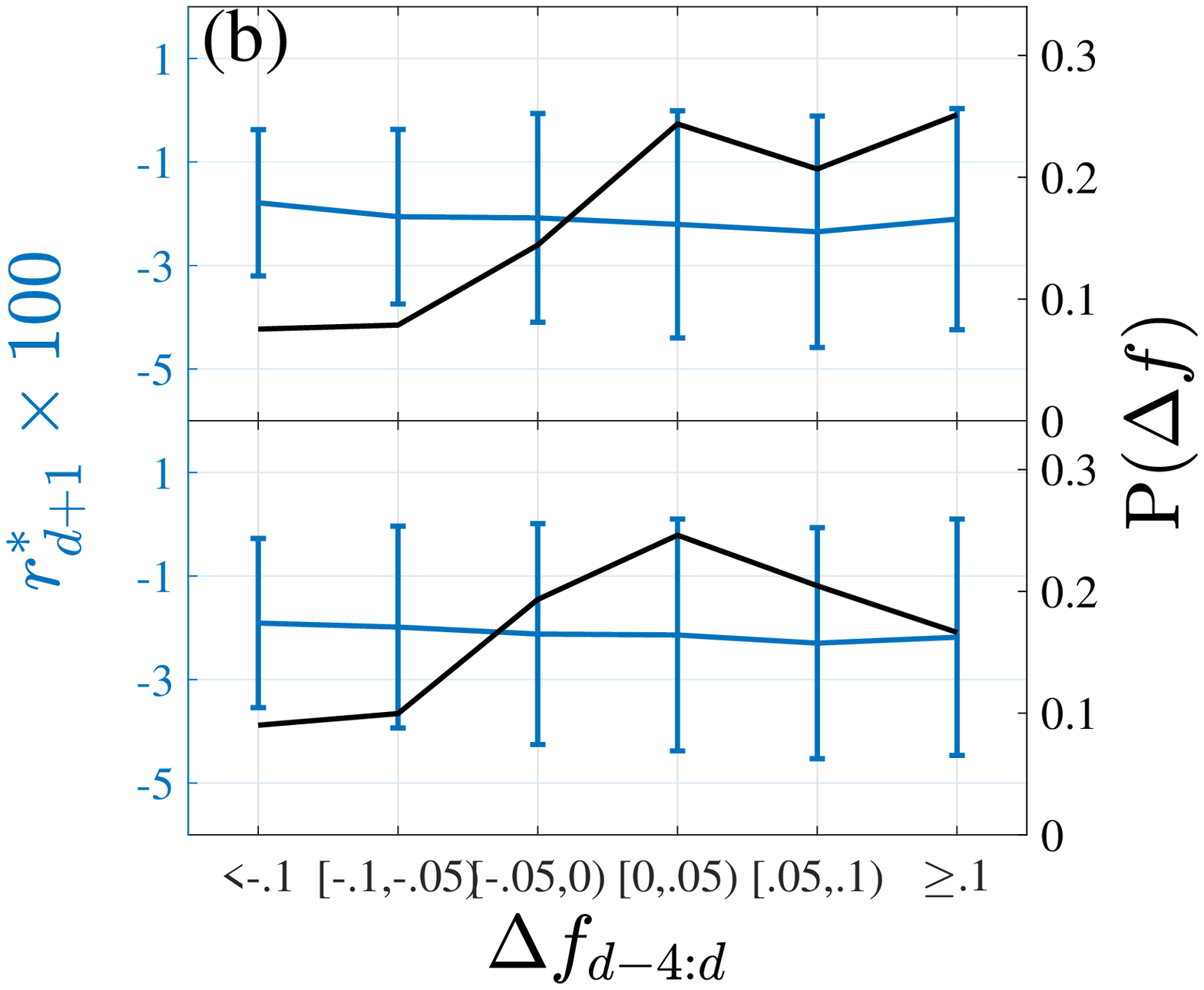} \hskip 3mm
\includegraphics[width=5cm]{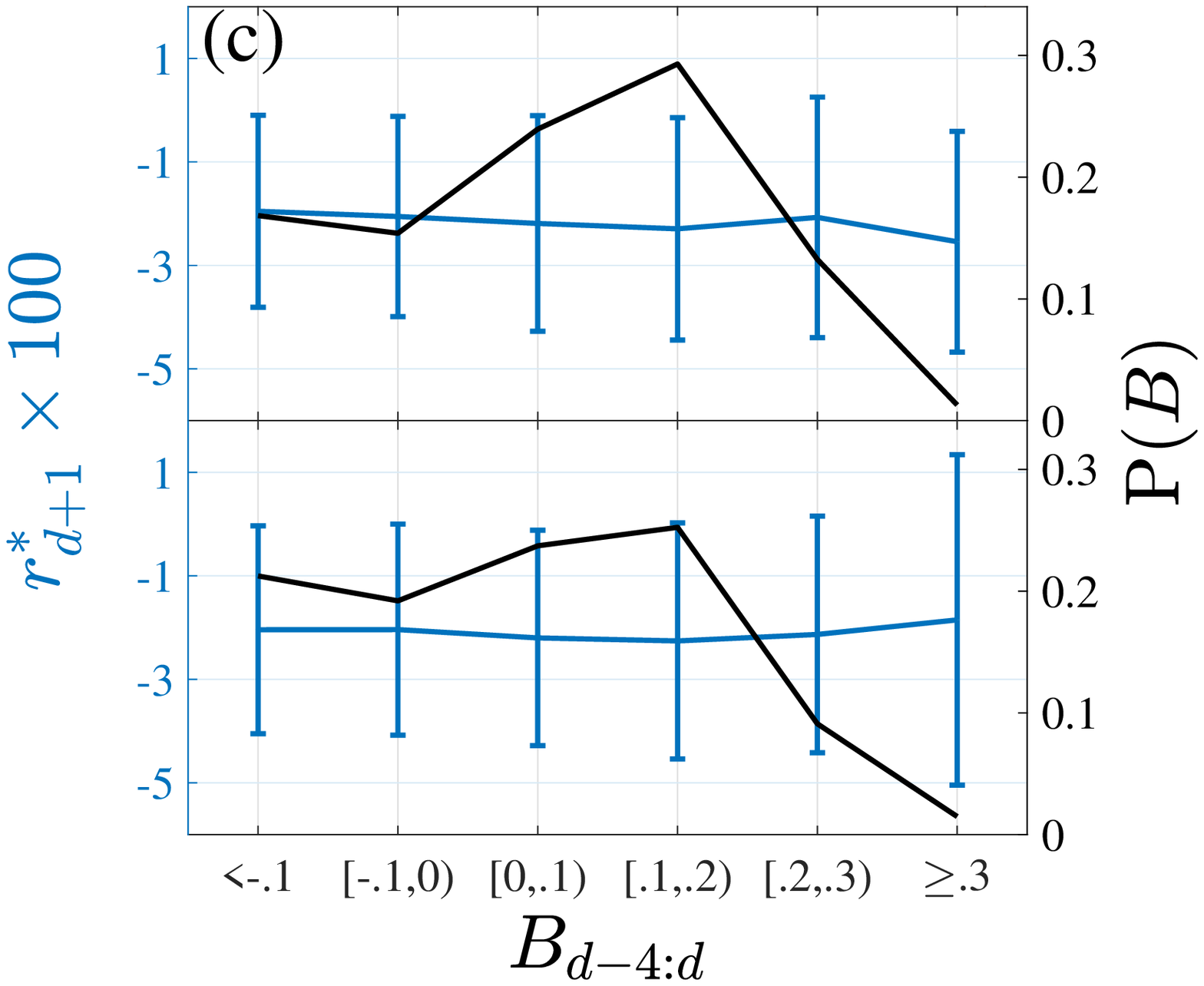}\hskip 3mm 
\caption{\label{Fig:MF:ER:MFC} Illustration of the predictive power of multifractal characteristics on the excess returns for SH 180 (top panels) and SZCI (bottom panels). (a) Plots of the average excess returns with respect to the average multifractal characteristic $\Delta {\alpha}$ on the left y-axis and the fraction of the multifractal characteristic $\Delta {\alpha}$ in each bin on the right y-axis. (b) The same as (a), but for the multifractal characteristic $\Delta f$. (c) The same as (a), but for the multifractal characteristic $B$. }
\end{figure}

The daily excess return $r^*$ on day $d$ is defined as the difference between the index return $r_d$ and the market risk-free return $r_f$, 
\begin{equation}
r^*(d) = r_d(d) - r_f(d).
\label{Eq:MFPredictReturn:ExcessReturn}
\end{equation}
And the corresponding multifractal characteristics on day $d$, denote as $d_{\alpha, d-4:d}$, $d_{f, d-4:d}$, and $B_{d-4:d}$, are estimated from day $(d-4)$ to day $d$. We first separate the multifractal characteristic $d_{\alpha, d-4:d}$ into six groups according to whether they fall into the following six bins, $(-\infty, 0.05]$, $(-0.05, 0.1]$, $(0.1, 0.15]$, $(0.15, 0.2]$, $(0.2, 0.25]$, and $(0.25, \infty)$. In each group, we calculate the average value of $d_{\alpha, d-4:d}$ and estimate the average value of excess returns $r^*_{d+1}$ which are exactly one day after $d_{\alpha, d-4:d}$. In Fig.~\ref{Fig:MF:ER:MFC} (a), we illustrate the errorbar plots of the average excess return $\langle r^*_{d+1} \rangle$ with respect to the average multifractal width $\langle d_{\alpha, d-4:d} \rangle$ on the left y-axis and the fraction of the multifractal width $d_{\alpha, d-4:d}$ in each bin on the right y-axis. The top panel is the results of SH180 and the bottom panel is the results of SZCI. One can see that there is a slightly increasing trend for $\langle r^*_{d+1} \rangle$ with the increasing of $\langle d_{\alpha, d-4:d} \rangle$ for both indexes, indicating the potential predictability of $\Delta {\alpha}$ on future excess returns. For the multifractal characteristic $\Delta f$ and $B$, we perform the same analysis as $\Delta {\alpha}$. The corresponding results are shown in Fig.~\ref{Fig:MF:ER:MFC} (b) and (c). One can observe a slightly decreasing trend for $\langle r^*_{d+1} \rangle$ with the increasing of $\langle d_{f, d-4:d} \rangle$ for both indexes, implying that there may exist to be a negative correlation between $\Delta f$ and the future excess returns. However, we cannot find any dependent behaviors between $\langle B \rangle$ and $r_d$, suggesting the uselessness of the multifractal characteristic $B$ on predicting future excess returns.

We further conduct the Granger causality tests between the excess returns $r^*$ and the multifractal characteristics $\Delta {\alpha}$, $d_{f}$, and $B$. The corresponding null hypothesis of  $x \nrightarrow y$ is that $x$ does not Granger cause $y$. The results are listed in Table~\ref{Tab:GrangerTests}. One can see that the excess return $r^*$ is the Granger causality of $\Delta {\alpha}$ and $B$ for SH180 at the level of 5\% and of $B$ for SZCI at the level of 1\%. The null hypothesis that $\Delta {\alpha}$ does not Granger cause $r^*$ is rejected for SH180 (respectively, SZCI) at the significant level of 0.1\% (5\%), indicating that the lagged $\Delta {\alpha}$ may contain the future information of the excess returns $r^*$. For the rest tests in Table \ref{Tab:GrangerTests}, the null hypothesis of the Granger tests cannot be rejected, meaning that there is no Granger causality between them. 

\begin{table}[htb]
\setlength\tabcolsep{16pt}
\centering %\small
\caption{\label{Tab:GrangerTests} Granger causality tests between the excess returns $r^*$and the multifractal characteristics, $\Delta {\alpha}$, $d_{f}$, and $B$. $x \nrightarrow y$ indicates the null hypothesis that $x$ does not Granger cause $y$. Note that $^{*}$, $^{**}$, and $^{***}$ indicates significant level of 5\%, 1\%, and 0.1\%, respectively.}
\begin{tabular}{cr@{.}lr@{.}lr@{.}lr@{.}lr@{.}lr@{.}l}
\toprule
Index & \multicolumn{2}{c}{$r^* \nrightarrow \Delta {\alpha}$} &  \multicolumn{2}{c}{$\Delta {\alpha} \nrightarrow r^*$} 
& \multicolumn{2}{c}{$r^* \nrightarrow d_{f}$} &  \multicolumn{2}{c}{$d_{f} \nrightarrow r^*$} 
& \multicolumn{2}{c}{$r^* \nrightarrow B$} &  \multicolumn{2}{c}{$B \nrightarrow r^*$} \\
\midrule
SH180 & 4&5297$^*$ & 17&4169$^{***}$ & 0&1289 & 0&3947 &  6&0398$^*$ & 0&0970\\
& (0&0334) & (0&0000) & (0&7196) & (0&5299) & (0&0140)& (0&7555)\\
SZCI & 1&1206 & 5&4089$^*$ &  0&5035 & 1&2445 & 7&9909$^{**}$ &  1&0561\\
& (0&2899) & (0&0201) & (0&4780) & (0&2647) & (0&0047) & (0&3042) \\
\bottomrule
\end{tabular}
\end{table}

A standard univariate predictive regression framework is employed to test the predictive power of mulitfractal characteristics $M_{d-4:d}$ from day ${d-4}$ to day $d$ on the excess returns on day $d+1$,
\begin{equation}
r^*_{d+1} = \alpha_r + \beta_{r, i} M_{i, d-4:d} + {\varepsilon _d}, d = 5, \cdots, T
\label{Eq:MFPredictReturn:ExcessReturn}
\end{equation}
where $M_1 = \Delta {\alpha}$, $M_2 = \Delta f$, $M_3 = B$, $M_4 = \Delta {\alpha} \Delta f$, $M_5 = \Delta {\alpha} B$, and $M_6 = \Delta f B$. The corresponding results of in-sample tests are shown in Table~\ref{Tab:FactorModel:InSampleTests} for both indexes, in which the regression slopes, intercepts, and the adjusted $R^2$ statistics are reported. The $p$-values, which are obtained from the NW $t$-tests \citep{Newey-West-1987-Em}, are also listed in the parentheses under the regression parameters. The null hypothesis of the NW $t$-tests is that the regressing slopes ($\beta_{r,i}$) are vanishing. If the null hypothesis is rejected for $\beta_{r,i}$, we can conclude that there are useful information for forecasting the future excess return $r^*_{d+1}$ in the predictor of multifractal characteristic $M_i$. And the acceptance of the null hypothesis means that the multifractal characteristic $M_i$ has no predictability in model (\ref{Eq:MFPredictReturn:ExcessReturn}). 

In panel A, we list the results of in-sample tests for the excess returns of SH180 index. Except $\Delta {\alpha}$, all the other multifractal characteristics give negative regression slopes. However, none of them is statistically significant. We can observe that only the regression slope of $\Delta {\alpha}$ is significant at the level of 0.1\%. As the excess return is hardly predictable, the adjusted $R^2$ statistics are all very small. The model based on $\Delta {\alpha}$ gives a $R^2$ statistics of 0.85\% for SH180, which means that the multifractal characteristic $\Delta {\alpha}$ can explain 0.85\% of the future excess returns. In panel B, we report the regressing parameters of in-sample tests for the excess returns of SZCI index. One can observe that the regressing slope of $\Delta {\alpha}$ (respectively, $\Delta f$) is positive (negative) and significant at the level of 1\% (5\%). And the corresponding $R^2$ statistics are 0.26\% and 0.13\% for $\Delta {\alpha}$ and $\Delta f$. The rest of the multifractal characteristics exhibit a weak predictability of the excess return, as their regression slopes are not statistically significant and adjust $R^2$ are quite small. Usually, the multifractal characteristic $\Delta {\alpha}$ can be regarded as a measurement of market volatilities. The predicting ability of $\Delta {\alpha}$ can be linked to the Chinese market risk. Higher market risk, associating with high market volatility, results in higher risk premium in the next period \citep{Merton-1980-JFE, French-Shwert-Stambaugh-1987-JFE}.

 %{\textcolor{red}{Comparing with the $R^2$ of other factors and citing more papers.}} 

\begin{table}[htbp]
\setlength\tabcolsep{7pt}
%\footnotesize
\centering %\small
\caption{\label{Tab:FactorModel:InSampleTests} In-sample tests of predicting the excess index return $r^*_{d+1}$ on day $d+1$ with multifractal characteristics $M_{d-4:d}$ from day $d-4$ to day $d$. The analysis are performed on the univariate predictive regression models: $r^*_{d+1} = \alpha_r + \beta_{r, i} M_{i, d-4:d} + {\varepsilon _d}$, where $M_1 = \Delta {\alpha}$, $M_2 = \Delta f,$, $M_3 = B$, $M_4 = \Delta {\alpha} \Delta f$, $M_5 = \Delta {\alpha} B$, and $M_6 = \Delta f B$. In each panel, we list the regression slope coefficients $\beta$, the regression intercepts $\alpha$, and the adjusted $R^2$ statistics. Note that $^{*}$, $^{**}$, and $^{***}$ indicates the significant level of 5\%, 1\%, and 0.1\%, respectively.}
\begin{tabular}{lr@{.}lr@{.}lr@{.}lr@{.}lr@{.}lr@{.}lr@{.}lr@{.}l}
\toprule
$M_{i, d-4:d}$ &  \multicolumn{2}{c}{$\beta_{r, 1}$} & \multicolumn{2}{c}{$\beta_{r, 2}$} & \multicolumn{2}{c}{$\beta_{r, 3}$} & \multicolumn{2}{c}{$\beta_{r, 4}$} & \multicolumn{2}{c}{$\beta_{r, 5}$} & \multicolumn{2}{c}{$\beta_{r, 6}$} & \multicolumn{2}{c}{$\alpha_r$} & \multicolumn{2}{c}{$R^2$} \\
\midrule
\multicolumn{17}{l}{Panel A: Regression results for SH180} \\
$\Delta {\alpha}$ & 0&0261$^{***}$ &  \multicolumn{2}{c}{} &  \multicolumn{2}{c}{} & \multicolumn{2}{c}{} & \multicolumn{2}{c}{} & \multicolumn{2}{c}{} & $-$0&0259$^{***}$ & 0&0085 \\
  & (0&0000) &  \multicolumn{2}{c}{} &  \multicolumn{2}{c}{} & \multicolumn{2}{c}{} & \multicolumn{2}{c}{} & \multicolumn{2}{c}{} & (0&0000) & \multicolumn{2}{c}{} \\

$\Delta f$ & \multicolumn{2}{c}{} &  $-$0&0041  &  \multicolumn{2}{c}{} & \multicolumn{2}{c}{} & \multicolumn{2}{c}{} & \multicolumn{2}{c}{} & $-$0&0213$^{***}$ & 0&0005 \\
  & \multicolumn{2}{c}{} & (0&2374)  &  \multicolumn{2}{c}{} & \multicolumn{2}{c}{} & \multicolumn{2}{c}{} & \multicolumn{2}{c}{} & (0&0000) & \multicolumn{2}{c}{} \\

$d_B$ & \multicolumn{2}{c}{} & \multicolumn{2}{c}{}  & $-$0&0001  & \multicolumn{2}{c}{} & \multicolumn{2}{c}{} & \multicolumn{2}{c}{} & $-$0&0215$^{***}$ & 0&0000  \\
  & \multicolumn{2}{c}{} & \multicolumn{2}{c}{}   & (0&8997)  & \multicolumn{2}{c}{} & \multicolumn{2}{c}{} & \multicolumn{2}{c}{} & (0&0000) & \multicolumn{2}{c}{} \\

$\Delta {\alpha}\Delta f$ & \multicolumn{2}{c}{} & \multicolumn{2}{c}{}  & \multicolumn{2}{c}{}   & $-$0&0020 & \multicolumn{2}{c}{} & \multicolumn{2}{c}{} & $-$0&0215$^{***}$ & 0&0000  \\
  & \multicolumn{2}{c}{} & \multicolumn{2}{c}{}   &  \multicolumn{2}{c}{}   &  (0&8841) & \multicolumn{2}{c}{} & \multicolumn{2}{c}{} & (0&0000) & \multicolumn{2}{c}{} \\
  
$\Delta {\alpha}B$ & \multicolumn{2}{c}{} & \multicolumn{2}{c}{}  & \multicolumn{2}{c}{}   & \multicolumn{2}{c}{}  &  $-$0&0220  & \multicolumn{2}{c}{} & $-$0&0215$^{***}$ & 0&0012  \\
  & \multicolumn{2}{c}{} & \multicolumn{2}{c}{}   &  \multicolumn{2}{c}{}   &  \multicolumn{2}{c}{} & (0&0552)   & \multicolumn{2}{c}{} & (0&0000) & \multicolumn{2}{c}{} \\

$\Delta fB$  & \multicolumn{2}{c}{} & \multicolumn{2}{c}{}  & \multicolumn{2}{c}{}   & \multicolumn{2}{c}{}  &  \multicolumn{2}{c}{}   & $-$0&0122  & $-$0&0216$^{***}$ & 0&0009 \\
  & \multicolumn{2}{c}{} & \multicolumn{2}{c}{}   &  \multicolumn{2}{c}{}   &  \multicolumn{2}{c}{} &  \multicolumn{2}{c}{}    & (0&1063) & (0&0000) & \multicolumn{2}{c}{} \\

\midrule
\multicolumn{17}{l}{Panel B: Regression results for SZCI} \\
$\Delta {\alpha}$ & 0&0169$^{**}$ &  \multicolumn{2}{c}{} &  \multicolumn{2}{c}{} & \multicolumn{2}{c}{} & \multicolumn{2}{c}{} & \multicolumn{2}{c}{} & $-$0&0240$^{***}$ & 0&0026 \\
  & (0&0048) &  \multicolumn{2}{c}{} &  \multicolumn{2}{c}{} & \multicolumn{2}{c}{} & 
\multicolumn{2}{c}{} & \multicolumn{2}{c}{} & (0&0000) & \multicolumn{2}{c}{} \\

$\Delta f$ & \multicolumn{2}{c}{} &  $-$0&0078$^*$  &  \multicolumn{2}{c}{} & \multicolumn{2}{c}{} & \multicolumn{2}{c}{} & \multicolumn{2}{c}{} & $-$0&0212$^{***}$ & 0&0013 \\
  & \multicolumn{2}{c}{} & (0&0467)  &  \multicolumn{2}{c}{} & \multicolumn{2}{c}{} & \multicolumn{2}{c}{} & \multicolumn{2}{c}{} & (0&0000) & \multicolumn{2}{c}{} \\

$d_B$ & \multicolumn{2}{c}{} & \multicolumn{2}{c}{}  & $-$0&0008  & \multicolumn{2}{c}{} & \multicolumn{2}{c}{} & \multicolumn{2}{c}{} & $-$0&0214$^{***}$ & 0&0003  \\
  & \multicolumn{2}{c}{} & \multicolumn{2}{c}{}   & (0&3709)  & \multicolumn{2}{c}{} & \multicolumn{2}{c}{} & \multicolumn{2}{c}{} & (0&0000) & \multicolumn{2}{c}{} \\
  
$\Delta {\alpha}\Delta f$ & \multicolumn{2}{c}{} & \multicolumn{2}{c}{}  & \multicolumn{2}{c}{}   & $-$0&0173 & \multicolumn{2}{c}{} & \multicolumn{2}{c}{} & $-$0&0213$^{***}$ & 0&0005  \\
  & \multicolumn{2}{c}{} & \multicolumn{2}{c}{}   &  \multicolumn{2}{c}{}   &  (0&2235) & \multicolumn{2}{c}{} & \multicolumn{2}{c}{} & (0&0000) & \multicolumn{2}{c}{} \\
  
$\Delta {\alpha}B$ & \multicolumn{2}{c}{} & \multicolumn{2}{c}{}  & \multicolumn{2}{c}{}   & \multicolumn{2}{c}{}  &  $-$0&0180  & \multicolumn{2}{c}{} & $-$0&0213$^{***}$ & 0&0006   \\
  & \multicolumn{2}{c}{} & \multicolumn{2}{c}{}   &  \multicolumn{2}{c}{}   &  \multicolumn{2}{c}{} & (0&1709)   & \multicolumn{2}{c}{} & (0&0000) & \multicolumn{2}{c}{} \\

$\Delta fB$  & \multicolumn{2}{c}{} & \multicolumn{2}{c}{}  & \multicolumn{2}{c}{}   & \multicolumn{2}{c}{}  &  \multicolumn{2}{c}{}   & $-$0&0007  & $-$0&0214$^{***}$ & 0&0000  \\
  & \multicolumn{2}{c}{} & \multicolumn{2}{c}{}   &  \multicolumn{2}{c}{}   &  \multicolumn{2}{c}{} &  \multicolumn{2}{c}{}    & (0&9384) & (0&0000) & \multicolumn{2}{c}{} \\
  
 \bottomrule
\end{tabular}
\end{table}

\subsection{Comparison with the Chinese market volatility measures}

To compare the predictive power of multifractal characteristics with the Chinese market volatility measures, we perform the following regression, 
\begin{equation}
r^*_{d+1} = \alpha_r +  \beta_{r, i} M_{i, d-4:d} + \psi_{r} v_{d-4:d} + {\varepsilon _d}, d = 5, \cdots, T
\label{Eq:MFPredictReturn:ExcessReturn:Volatility}
\end{equation}
where $v_{d-4:d}$ is the realized market volatility from day $d-4$ to $d$, which is estimated by summing the square of minutely returns in sample periods. 

Table.~\ref{Tb:RetPre:Comparison:Volatility} reports the results of the regression to Eq.~(\ref{Eq:MFPredictReturn:ExcessReturn:Volatility}). We find that the results in Table.~\ref{Tb:RetPre:Comparison:Volatility} are in accordance with those in Table.~\ref{Tab:FactorModel:InSampleTests} and that only the $\beta$ of $\Delta {\alpha}$ is statistically significant for both indexes and the other predictors including the realized volatility are economically insignificant, which reveals that the multifractal characteristic $\Delta {\alpha}$ do have additional information beyond that of market volatility. 

\begin{table*}[tb]
\setlength\tabcolsep{4pt}
%\footnotesize
\centering %\small
 \caption{\label{Tb:RetPre:Comparison:Volatility} Comparison of the predictive ability between the multifractal characteristic and the realized volatility. The analysis are performed on the bivariate predictive regression models: $r^*_{d+1} = \alpha_r +  \beta_{r, i} M_{i, d-4:d} + \psi_{r} v_{d-4:d} + {\varepsilon _d}$, where $M_1 = \Delta {\alpha}$, $M_2 = \Delta f$, $M_3 = B$, $M_4 = \Delta {\alpha} \Delta f$, $M_5 = \Delta {\alpha} B$, and $M_6 = \Delta f B$. In each panel, we list the regression slope coefficients $\beta$ and $\psi$, the regression intercepts $\alpha$, and the adjusted $R^2$ statistics. Note that $^{*}$, $^{**}$, and $^{***}$ indicates the significant level of 5\%, 1\%, and 0.1\%, respectively. }
  \medskip
 \centering
\begin{tabular}{lcr@{.}lr@{.}lr@{.}lr@{.}lr@{.}lr@{.}lr@{.}lr@{.}lr@{.}l}
\toprule
  $M_{i, d-4:d}$ &&\multicolumn{2}{c}{$\psi_{r}$} & \multicolumn{2}{c}{$\beta_{r, 1}$} & \multicolumn{2}{c}{$\beta_{r, 2}$} & \multicolumn{2}{c}{$\beta_{r, 3}$} & \multicolumn{2}{c}{$\beta_{r, 4}$} & \multicolumn{2}{c}{$\beta_{r, 5}$} & \multicolumn{2}{c}{$\beta_{r, 6}$} & \multicolumn{2}{c}{$\alpha_r$} & \multicolumn{2}{c}{$R^2$} \\
  \midrule
  \multicolumn{19}{l}{Panel A: Regression results for SH180} \\
  $\Delta {\alpha}$ && 0&1727 & 0&0265***  &\multicolumn{2}{c}{}  &\multicolumn{2}{c}{}  &\multicolumn{2}{c}{}  &\multicolumn{2}{c}{}  &\multicolumn{2}{c}{} & $-$0&0262*** & 0&0087 \\
  && (0&4046) & (0&0000)  &\multicolumn{2}{c}{}  &\multicolumn{2}{c}{}  &\multicolumn{2}{c}{}  &\multicolumn{2}{c}{}  &\multicolumn{2}{c}{} & (0&0000) &\multicolumn{2}{c}{} \\
  $d_{f}$ && 0&1183  &\multicolumn{2}{c}{} & $-$0&0045   &\multicolumn{2}{c}{}  &\multicolumn{2}{c}{}  &\multicolumn{2}{c}{}  &\multicolumn{2}{c}{} & $-$0&0214*** & 0&0006 \\
  && (0&5744)   &\multicolumn{2}{c}{} & (0&2065)   &\multicolumn{2}{c}{}  &\multicolumn{2}{c}{}  &\multicolumn{2}{c}{}  &\multicolumn{2}{c}{} &(0&0000) &\multicolumn{2}{c}{}  \\
  $B$ && 0&0713  &\multicolumn{2}{c}{}  &\multicolumn{2}{c}{} & $-$0&0001 &  \multicolumn{2}{c}{}  &\multicolumn{2}{c}{}  &\multicolumn{2}{c}{} & $-$0&0215*** & 0&0000 \\
 & & (0&7308)  &\multicolumn{2}{c}{}  &\multicolumn{2}{c}{}  & (0&8910) &  \multicolumn{2}{c}{}  &\multicolumn{2}{c}{}  & \multicolumn{2}{c}{} & (0&0000) &\multicolumn{2}{c}{}  \\
  $\Delta {\alpha}*d_{f}$ &\multicolumn{2}{c}{} 0&0671  &\multicolumn{2}{c}{}  &\multicolumn{2}{c}{} &\multicolumn{2}{c}{} & 0&0011  &\multicolumn{2}{c}{}  &\multicolumn{2}{c}{} & $-$0&0216*** & 0&0000 \\
 && (0&7508)   &\multicolumn{2}{c}{}  &\multicolumn{2}{c}{}  &\multicolumn{2}{c}{} & (0&9352)  &\multicolumn{2}{c}{}  &\multicolumn{2}{c}{}  & (0&0000) &\multicolumn{2}{c}{}  \\
  $\Delta {\alpha}*B$ && 0&1031 &  \multicolumn{2}{c}{}  &\multicolumn{2}{c}{}  &\multicolumn{2}{c}{}  &\multicolumn{2}{c}{} &  $-$0&0225 &  \multicolumn{2}{c}{} & $-$0&0214*** & 0&0013 \\
  && (0&6197) &  \multicolumn{2}{c}{} &  \multicolumn{2}{c}{} &  \multicolumn{2}{c}{} &  \multicolumn{2}{c}{} & (0&0511) & \multicolumn{2}{c}{}  & (0&0000) &\multicolumn{2}{c}{}  \\
  $d_{f}*B$ && 0&0805 &  \multicolumn{2}{c}{} &  \multicolumn{2}{c}{} &  \multicolumn{2}{c}{} &  \multicolumn{2}{c}{} &  \multicolumn{2}{c}{} & 0&0122 &$ -$0&0217*** & 0&0009 \\
  && (0&6977) &  \multicolumn{2}{c}{} &  \multicolumn{2}{c}{} &  \multicolumn{2}{c}{} &  \multicolumn{2}{c}{} &  \multicolumn{2}{c}{}& (0&1040) & (0&0000) &\multicolumn{2}{c}{}  \\
  \midrule
  \multicolumn{19}{l}{ Panel B: Regression results for SZCI } \\
  $\Delta {\alpha}$ && $-$0&0546 & 0&0168** &  \multicolumn{2}{c}{} &  \multicolumn{2}{c}{} &  \multicolumn{2}{c}{} &  \multicolumn{2}{c}{} &  \multicolumn{2}{c}{} & $-$0&0239*** & 0&0026 \\
  && (0&8315) & (0&0051) &  \multicolumn{2}{c}{} &  \multicolumn{2}{c}{} &  \multicolumn{2}{c}{} &  \multicolumn{2}{c}{} &  \multicolumn{2}{c}{} & (0&0000) &\multicolumn{2}{c}{}  \\
  $d_{f}$ && 0&0050 &   \multicolumn{2}{c}{} & $-$0&0078 &  \multicolumn{2}{c}{} &  \multicolumn{2}{c}{} &  \multicolumn{2}{c}{} &  \multicolumn{2}{c}{} & $-$0&0212*** & 0&0013 \\
  && (0&9849) &   \multicolumn{2}{c}{} & (0&0511) &  \multicolumn{2}{c}{} &  \multicolumn{2}{c}{} &  \multicolumn{2}{c}{} &  \multicolumn{2}{c}{} & (0&0000) &\multicolumn{2}{c}{}   \\
  $B$ && $-$0&0868 &  \multicolumn{2}{c}{} &  \multicolumn{2}{c}{} & $-$0&0008 &  \multicolumn{2}{c}{} &  \multicolumn{2}{c}{} &  \multicolumn{2}{c}{} & $-$0&0213*** & 0&0003 \\
  && (0&7355) &  \multicolumn{2}{c}{} &  \multicolumn{2}{c}{} & (0&3822) &  \multicolumn{2}{c}{} &  \multicolumn{2}{c}{} &  \multicolumn{2}{c}{} & (0&0000) &\multicolumn{2}{c}{}   \\
  $\Delta {\alpha}*d_{f}$ && $-$0&0179 &  \multicolumn{2}{c}{} &  \multicolumn{2}{c}{} &  \multicolumn{2}{c}{} & $-$0&0171 &  \multicolumn{2}{c}{} &  \multicolumn{2}{c}{} & $-$0&0213*** & 0&0005 \\
  && (0&9461) &  \multicolumn{2}{c}{} &  \multicolumn{2}{c}{} &  \multicolumn{2}{c}{} & (0&2479) &  \multicolumn{2}{c}{} &  \multicolumn{2}{c}{} & (0&0000) &\multicolumn{2}{c}{}   \\
  $\Delta {\alpha}*B$ && $-$0&0513 &  \multicolumn{2}{c}{} &  \multicolumn{2}{c}{} &  \multicolumn{2}{c}{} &  \multicolumn{2}{c}{} & $-$0&0176 &  \multicolumn{2}{c}{} & $-$0&0213*** & 0&0006 \\
  && (0&8430) &  \multicolumn{2}{c}{} &  \multicolumn{2}{c}{} &  \multicolumn{2}{c}{} &  \multicolumn{2}{c}{} & (0&1843) &  \multicolumn{2}{c}{} & (0&0000) &\multicolumn{2}{c}{}   \\
  $d_{f}*B$ && $-$0&0991 &  \multicolumn{2}{c}{} &  \multicolumn{2}{c}{} &  \multicolumn{2}{c}{} &  \multicolumn{2}{c}{} &  \multicolumn{2}{c}{} & 0&0006 & $-$0&0213*** & 0&0001 \\
  && (0&6993) &  \multicolumn{2}{c}{} &  \multicolumn{2}{c}{} &  \multicolumn{2}{c}{} &  \multicolumn{2}{c}{} &  \multicolumn{2}{c}{} & (0&9486) & (0&0000) &\multicolumn{2}{c}{}   \\
\bottomrule
\end{tabular}
\end{table*}

\subsection{Out-of-sample tests}

Out-of-sample tests are usually encouraged to evaluate the return predictability by excluding the using of future information and over-fitting in-sample tests. Two statistics, the $R_{OS}^2$ statistic \citep{Campbell-Thompson-2008-RFS} and adjusted MSFE statistic \citep{Clark-West-2008-JEm}, are employed to assess the out-of-sample forecasting performance. The $R_{OS}^2$ statistic is defined as follows, 
\begin{equation}
 R_{OS}^2 = 1 - \frac{\sum_{d=n}^{T-1} \left( r_{d+1}^* - \hat{r}^*_{d+1)} \right)}{ \sum_{d=n}^{T-1} \left( r_{d+1}^* - \bar{r}_{d+1}^* \right) }
  \label{Eq:scalingfunction}
\end{equation}
where $r_{d+1}^*$ is the actual excess return, $\hat{r}_{d+1}^*$ is the predicting excess return, and $\bar{r}_{d+1}^*$ is the benchmark of historical average returns. $R_{OS}^2$ measures the percent reduction in mean square forecast error for the predictive regression forecast relative to the historical average benchmark forecast. From the definition, we can infer that $R_{OS}^{2}$ locates in the range of $(\infty, 1]$. The predictive regression forecast $\hat{r}_{d+1}^*$ can be considered better than the historical average $\bar{r}_{d+1}^*$ from the perspective of mean squared forecasting errors (MSFE) when $R_{OS}^2 > 0$. The adjusted MSFE statistic, proposed by \cite{Clark-West-2008-JEm}, is used to indicate the statistical significance of $R_{OS}^2 > 0$ and is included in \cite{Neely-Rapach-Tu-Zhou-2014-MS}, \cite{Phan-Sharma-Narayan-2015-IRFA}, \cite{Guo-Tao-2017-SSRN}, \cite{Chen-Jiang-Liu-Tu-2017-JIMF}, and so on. The null hypothesis of the adjusted MSFE statistic is $R_{OS}^2 \le 0$, corresponding to that the MSFE of the historical average benchmark is less than or equal to that of predictive regression forecast.

Our out-of-sample tests are conducted in both expanding windows and moving windows. The detailed procedure is listed as follows. The predictive regression (Eq.~(\ref{Eq:MFPredictReturn:ExcessReturn})) is estimated in each window and the obtained parameters are then used to generate the out-of-sample prediction for the day following that window. These steps are repeated by expanding or sliding the window till one reach the end of sample period. The initial window includes the first 600 data points, about 2 years of data. The results are listed in Table.~\ref{Tb:RetPre:OutofSample:Tests}. Panel A lists the results of the out-of-sample tests for SH180, one can find that only $\Delta {\alpha}$ and $\Delta f$ have positive $R_{\rm{OS}}^2$ in both moving windows and expanding windows, indicating that the predictive regression forecast is superior to the historical average, and the rest predictors all have  $R_{\rm{OS}}^2 < 0$. According to the adjusted MSFE statistics, we find that the $R_{OS}^2$ not greater than 0 is rejected for $\Delta {\alpha}$ and $\Delta f$ at the significant level of 0.1\% and 5\% in the moving window and for $\Delta {\alpha}$ at the significant level of 1\% in the expanding window. The results of the out-of-sample tests for SZCI are reported in Panel B of Table.~\ref{Tb:RetPre:OutofSample:Tests}. One can see that $R_{OS}^2$ corresponding to predictors $d_B$ and $\Delta {\alpha} B$ are negative and other predictors all have positive $R_{OS}^2$ in the moving window and expanding window. However, only half of them are statistically significant according to the adjusted MSFE statistics, such as the predictors $\Delta {\alpha}$, $\Delta f$, $\Delta fB$ in the moving window tests and the predictor $\Delta f$ in the expanding window tests. We also find that including more data points into the predictive regression forecast will shrink the predictive ability of the predictor, evidenced by lower $R_{OS}^2$ in expanding window tests comparing with that in moving window tests. The out-of-sample tests also reveal that the multifractal characteristics $\Delta {\alpha}$ and $\Delta f$ are good return predictors. 

\begin{table}[htb]
\setlength\tabcolsep{2.5pt}
%\footnotesize
\centering %\small
 \caption{\label{Tb:RetPre:OutofSample:Tests} Out-of-sample tests of predicting the excess index return $r^*_{d+1}$ on day $d+1$ with multifractal characteristics $M_{d-4:d}$ from day $d-4$ to day $d$. The $R_{OS}^2$ and adjusted MSFE statistics are listed. $R_{OS}^2$ measures the percent reduction in mean square forecast error for the predictive regression forecast based on the multifractal characteristic given in the first column relative to the historical average benchmark forecast. The adjusted MSFE statistic evaluates the significant level of  $R_{OS}^2 >0$.  Note that $^{*}$, $^{**}$, and $^{***}$ indicates the significant level of 5\%, 1\%, and 0.1\%, respectively. }
 \medskip
 \centering
\begin{tabular}{lcr@{.}lr@{.}lcr@{.}lr@{.}lcr@{.}lr@{.}lcr@{.}lr@{.}l}
\toprule
&& \multicolumn{9}{c}{Panel A: SH180} && \multicolumn{9}{c}{Panel B: SZCI}\\  %
 \cline{3-11} \cline{13-21}
 && \multicolumn{4}{c}{Moving window} && \multicolumn{4}{c}{Expanding window} && 
 \multicolumn{4}{c}{Moving window} && \multicolumn{4}{c}{Expanding window} \\
  \cline{3-6} \cline{8-11} \cline{13-16} \cline{18-21} 
&&  \multicolumn{2}{c}{$R_{OS}^2$} & \multicolumn{2}{c}{MSFE\_adj} && \multicolumn{2}{c}{$R_{OS}^2$}  & \multicolumn{2}{c}{MSFE\_adj}  &&  \multicolumn{2}{c}{$R_{OS}^2$}  &  \multicolumn{2}{c}{MSFE\_adj} && \multicolumn{2}{c}{$R_{OS}^2$}  & \multicolumn{2}{c}{MSFE\_adj} \\
\midrule
$\Delta {\alpha}$ && 0&88\% &  3&63***  && 0&65\%  &  3&10** &&  0&59\% & 3&03** && 0&05\%& 0&99  \\
$d_{f}$ && 0&52\% &  2&56*  && 0&16\%  &  1&63 &&  0&92\% & 3&60*** && 0&42\%  & 2&59** \\
$d_{B}$ &&  $-$1&22\% &  1&37  && $-$0&63\%  &  0&34 && $-$1&12\% &  0&64 && $-$0&25\% & 0&45 \\
$\Delta {\alpha}\Delta f$ && $-$0&03\% & 0&75  && $-$0&00\%  &  0&54 &&  0&28\% & 1&80 && 0&11\% & 1&34 \\
$\Delta {\alpha}B$ && $-$0&49\% &  $-$0&24  && $-$0&26\%  &  0&45 &&  $-$0&03\% & 1&24 && $-$0&07\% & 0&49 \\
$\Delta fB$ && $-$0&23\% &  1&23  && $-$0&11\%  & 0&31 &&  0&17\% & 2&29*&& 0&01\% & 1&50  \\
  
 \bottomrule
\end{tabular}
\end{table}

\section{Conclusion}
\label{Sec:Conclusion}

In this paper, we apply MF-DFA to detect the multifractal characteristics in high-frequency data of SH180 Index and SZSE Index. We find that both indexes exhibit strong multifractality. We propose three volatility measures based on the multifractal spectra ($\Delta {\alpha}$, $\Delta f$, and $B$) obtained from returns in moving windows with a size of five days. It is found that there is a positive correlation between $\Delta {\alpha}$ and short-term future returns, and a negative correlation between $\Delta f$ and short-term future returns. With the factor model, the return predictability is observed for the multifractal characteristic of spectral width $\Delta {\alpha}$ in both in-sample and out-of-sample tests, such that larger spectral widths result in higher excess returns. 

One possible explanation of such predicting ability is that multifractal characteristics are considered as measures of market volatility and can be linked to market risk, the return predictability may be economically explained by the theory of risk premium \citep{Merton-1980-JFE, French-Shwert-Stambaugh-1987-JFE}. Another possible explanation is that stock returns do not exhibit linear long memory, but possess nonlinear long memory and the multifractality is able to capture such nonlinear long memory, which may contain the information of future returns. The following work can be focused on  investigating the economic meaning of the predicting ability for multifractal characteristics, testing the return predictability of multifractal characteristics on various industry sectors and regions, and designing trading rules based on multifractal characteristics.

\bigskip
{\textbf{Acknowledgments:}}

This work was partially supported by the National Natural Science Foundation of China (71571121),  the Shanghai Philosophy and Social Science Fund Project (2017BJB006) and the Fundamental Research Funds for the Central Universities (222201718006). 

%\bibliography{E:/Papers/Auxiliary/Bibliography}

%\bibliography{/Users/zqjiang/research/Papers/Auxiliary/Bibliography}

\end{document}